\documentclass[12pt]{article}
\usepackage{epsfig}
\usepackage{array}
\usepackage{float}
\usepackage{lscape,graphicx}
\usepackage{graphics}
\usepackage{amssymb}
\usepackage{color}

\def\ltap{\ \raisebox{-.4ex}{\rlap{$\sim$}} \raisebox{.4ex}{$<$}\ }
\def\gtap{\ \raisebox{-.4ex}{\rlap{$\sim$}} \raisebox{.4ex}{$>$}\ }

\newcommand{\bea}{\begin{equation}\begin{array}{c}}
\newcommand{\eea}{\end{array}\end{equation}}
\newcommand{\ea}{\end{array}}

\newcommand{\beq}{\begin{equation}}
\newcommand{\eeq}{\end{equation}}
\newcommand{\bad}{\begin{array}{ccc}}

\newcommand{\ba}{\begin{array}{c}}
\newcommand{\half}{\frac{1}{2}}
\newcommand{\diag}{{\rm diag}}

\begin{document}
\hfill{{\small Ref. SISSA 25/2013/FISI}}

\begin{center}
\mathversion{bold}
{\bf{\large Lepton Flavor Violating  $\tau$ Decays in TeV Scale
Type I See-Saw and Higgs Triplet Models}}
\mathversion{normal}

\vspace{0.4cm}
D. N. Dinh$\mbox{}^{a,b)}$,
S. T. Petcov$\mbox{}^{a,c)}$~
\footnote{Also at: Institute of Nuclear Research and
Nuclear Energy, Bulgarian Academy of Sciences, 1784 Sofia, Bulgaria}

\vspace{0.2cm}
$\mbox{}^{a)}${\em  SISSA and INFN-Sezione di Trieste, \\
Via Bonomea 265, 34136 Trieste, Italy.\\}

\vspace{0.1cm}
$\mbox{}^{b)}${\em Institute of Physics, Vietnam Academy
of Science and Technology, \\
10 Dao Tan, Hanoi, Vietnam.\\
}

\vspace{0.1cm}
$\mbox{}^{c)}${\em Kavli IPMU, University of Tokyo (WPI), Tokyo, Japan.\\
}
\end{center}

\begin{abstract}
The lepton flavour violating (LFV) $\tau$ decays $\tau\to (e,\mu)\gamma$
and $\tau\to 3\mu$ are investigated in the frameworks of the
TeV scale type I see-saw and Higgs Triplet (or type II see-saw) models.
Predictions for the rates of these processes are obtained.
The implications of the existing stringent experimental upper bounds
on the $\mu\to e + \gamma$ and $\mu\to 3e$ decay branching ratios
for the predictions of the $\tau\to (e,\mu)\gamma$
and $\tau\to 3\mu$ decay rates are studied in detail.
The possibilities to observe the indicated LFV $\tau$ decays
in present and future experiments are analysed.
\end{abstract}

\section{Introduction}

 It is well established at present
that the flavour neutrino oscillations
observed in the experiments
with solar, atmospheric, reactor and accelerator neutrinos
(see~\cite{PDG2012} and the references quoted therein)
are caused by the existence of
nontrivial neutrino mixing in the weak
charged current interaction Lagrangian:
\begin{equation}
\label{eq:CC}
{\cal L}_{\rm CC} = - ~\frac{g}{\sqrt{2}}\,
\sum_{l=e,\mu,\tau}
\overline{l_L}(x)\, \gamma_{\alpha} \nu_{lL}(x)\,
W^{\alpha \dagger}(x) + h.c.\,,~~
\nu_{l \mathrm{L}}(x)  = \sum^n_{j=1} U_{l j} \nu_{j \mathrm{L}}(x),
\end{equation}
\noindent where
$\nu_{lL}(x)$ are the flavour neutrino fields,
$\nu_{j \mathrm{L}}(x)$ is the left-handed (LH)
component of the field of
the neutrino $\nu_j$ having a mass $m_j$,
and $U$ is a unitary matrix - the
Pontecorvo-Maki-Nakagawa-Sakata (PMNS)
neutrino mixing matrix~\cite{BPont67,BPont57,MNS62},
$U\equiv U_{PMNS}$.
The data imply that among the neutrinos
with definite mass at least three,
say $\nu_1$, $\nu_2$ and $\nu_3$,
have masses $m_{1,2,3} \lesssim 1$ eV, i.e.,
are much lighter than the charged leptons and quarks.

  The mixing of the three light neutrinos is described
to a good approximation by $3\times3$ unitary PMNS matrix.
In the widely used standard parametrisation \cite{PDG2012},
$U_{\rm PMNS}$ is expressed in terms of the solar,
atmospheric and reactor neutrino mixing angles
$\theta_{12}$,  $\theta_{23}$ and
$\theta_{13}$, respectively, and one Dirac - $\delta$, and
two Majorana \cite{BHP80} - $\alpha_{21}$ and $\alpha_{31}$,
CP violation (CPV) phases:
\begin{equation}
U_{\rm PMNS} = \tilde{U} P\,,~~~
P = {\rm diag}(1, e^{i \frac{\alpha_{21}}{2}}, e^{i \frac{\alpha_{31}}{2}})\,,
\label{01-VP}
\end{equation}
%
where
$\tilde{U}$ is a CKM-like matrix
containing the Dirac CPV
phase $\delta$,
\begin{equation}
\begin{array}{c}
\label{01-eq:Vpara}
\tilde{U} = \left(\begin{array}{ccc} 
 c_{12} c_{13} & s_{12} c_{13} & s_{13} e^{-i \delta}  \\[0.2cm]
 -s_{12} c_{23} - c_{12} s_{23} s_{13} e^{i \delta}
 & c_{12} c_{23} - s_{12} s_{23} s_{13} e^{i \delta}
 & s_{23} c_{13} 
\\[0.2cm]
 s_{12} s_{23} - c_{12} c_{23} s_{13} e^{i \delta} &
 - c_{12} s_{23} - s_{12} c_{23} s_{13} e^{i \delta}
 & c_{23} c_{13} 
\\
  \end{array} 
\right)\,.
\end{array}
\end{equation}
%
\noindent
In eq.~(\ref{01-eq:Vpara})
we have used the standard notations
$c_{ij} = \cos\theta_{ij}$, $s_{ij} = \sin\theta_{ij}$,
the angles $\theta_{ij} = [0,\pi/2]$,
$\delta = [0,2\pi]$
and, in general,
$0\leq \alpha_{j1}/2\leq 2\pi$, $j=2,3$~
\cite{EMSPEJP09}.
If CP invariance holds, we have
$\delta =0,\pi$, and
\cite{LW81}
$\alpha_{21(31)} = k^{(')}\,\pi$, $k^{(')}=0,1,2,3,4$.

The neutrino oscillation data, accumulated over many years,
allowed to determine the parameters which drive the
observed solar, reactor, atmospheric and accelerator
neutrino oscillations, $\Delta m^{2}_{21} > 0$, $\theta_{12}$,
$|\Delta m^{2}_{31}| \cong |\Delta m^{2}_{32}|$, $\theta_{23}$
and $\theta_{13}$ with a relatively
high precision (see, e.g., \cite{PDG2012}).
%
%
The best fit values of
these parameters, obtained in
\cite{Fogli:2012XY} from fitting the
global neutrino oscillation data, read:
\begin{eqnarray}
\label{Deltam2osc}
\Delta m^2_{21} = 7.54 \times 10^{-5} \ {\rm eV^2}\,,
|\Delta m^2_{31(32)}| = 2.47~(2.46) \times 10^{-3} \ {\rm eV^2}\,,\\
\sin^2\theta_{12} = 0.307\,,~
\sin^2\theta_{13} = 0.0241~(0.0244)\,,~\sin^2\theta_{23} = 0.386~(0.392)\,,
\label{thetaosc}
\end{eqnarray}
%
where the values (values in brackets) correspond to $\Delta m^2_{31(32)} > 0$
($\Delta m^2_{31(32)} < 0$), i.e., to
neutrino mass spectrum with normal ordering (NO),
$m_1 < m_2 < m_3$ (inverted ordering (IO), $m_3 < m_1 < m_2$)~
\footnote{As is well known, depending on the value of
the lightest neutrino mass, the spectrum can also be
normal hierarchical (NH), $m_1 \ll m_2 < m_3$;
inverted hierarchical (IH): $m_3 \ll m_1 < m_2$;
or quasi-degenerate (QD): $m_1\cong m_2\cong m_3$,
$m^2_j \gg |\Delta m^2_{31(32)}|$.
The QD spectrum corresponds to
$m_j \gtap 0.10$ eV, $j=1,2,3$.
}
(see, e.g., \cite{PDG2012}).
We will use these values in our numerical analyses.
Similar results have been obtained also in the global
analysis of the neutrino oscillation data performed in
\cite{Gonzalez-Garcia:2012}.

 In spite of the compelling evidence for nonconservation
of the leptonic flavour in neutrino oscillations,
reflected in the neutrino mixing present in eq. (\ref{eq:CC}),
all searches for lepton flavour violation (LFV) in the charged lepton sector
have produced negative results so far. The most stringent
upper limits follow from the experimental searches for the
LFV muon decays $\mu^+\rightarrow e^+\gamma$ and
$\mu^+\rightarrow e^+ e^- e^+$,
\begin{eqnarray}
\label{mutoegexp}
&~&{\rm BR}(\mu^+\rightarrow e^+\gamma)
< 5.7\times 10^{-13}~~~\mbox{\cite{MEG032013}}\,,\\
&~&{\rm BR}(\mu^+\rightarrow e^+ e^- e^+)< 1.0\times 10^{-12}~~
 \mbox{\cite{Bellgardt:1987du}}\,,
\label{muto3eexp}
\end{eqnarray}
%
and from the non-observation of conversion of muons
into electrons in Titanium,
\begin{equation}
{\rm CR}(\mu^- + {\rm Ti}\rightarrow e^- + {\rm Ti})\, <\,
4.3\times 10^{-12}~~~\mbox{\cite{Dohmen:1993mp}}\,.
\label{mu2eTi}
\end{equation}
%
  Besides, there are stringent constraints on the tau-muon
and tau-electron flavour violation as well from the non-observation of
LFV tau decays ~\cite{Aubert:2009tk}:
\begin{eqnarray}
\label{tautomugexp}
&~&{\rm BR}(\tau\rightarrow \mu\gamma)<4.4\times 10^{-8}\,,\\
\label{tautoegexp}
&~&{\rm BR}(\tau\rightarrow e \gamma)<3.3\times 10^{-8}\,,\\
\label{tauto3muexp}
&~&{\rm BR}(\tau\rightarrow 3\mu)<2.1\times 10^{-8}\,.
\end{eqnarray}
%

 In the minimal extension of the Standard Model
with massive neutrinos \cite{Petcov:1976ff},
in which the total lepton charge $L$ is conserved
($L = L_e + L_\mu + L_\tau$
$L_l$, $l=e,\mu,\tau$, being the individual lepton charges)
and the neutrinos with definite mass
are Dirac particles, the rates of the LFV
violating processes involving the charged leptons are
extremely strongly suppressed, which makes them
unobservable in practice. Indeed,
the $\mu\rightarrow e +\gamma$
decay branching ratio, for instance,
is given by \cite{Petcov:1976ff}:
\beq
BR(\mu \rightarrow e + \gamma) = \frac{3\alpha}{32\pi}\,
\left | U_{ej}\,U^{*}_{\mu j}\, \frac{m^2_j}{M^2_W} \right |^2
\cong (2.5 - 3.9)\times 10^{-55}\,,
\label{muegDnus}
\eeq
%
where we have used the best fit values of
the neutrino oscillation
parameters given in eqs. (\ref{Deltam2osc}) and (\ref{thetaosc})
and the two values of  $BR(\mu \rightarrow e + \gamma)$
correspond to $\delta = \pi$ and 0. The
predicted branching ratio should be compared with
the current experimental upper limit
quoted in eq. (\ref{mutoegexp}).

  The minimal extension of the Standard Model with
massive Dirac neutrinos and conservation of the
total lepton charge $L$ \cite{Petcov:1976ff}
does not give us, however,  any insight of
why the neutrino masses are so much smaller
than the masses of the charged leptons and quarks.
The enormous disparity between the magnitude of
the neutrino masses and the masses of
the charged fermions
suggests that the neutrino masses are related to the existence
of new mass scale $\Lambda$ in physics, i.e., to new physics
beyond the Standard Model (SM). A natural explanation of
the indicated disparity is provided by the see-saw
models of neutrino mass generation.
In the present study we will be primarily
interested in the type I seesaw \cite{seesaw} and
the Higgs Triplet (HT) \cite{typeII}
scenarios. In these models the scale $\Lambda$ is set by the
scale of masses of the new degrees of freedom
present in the models. In the case of the
type I see-saw scenario, these are the masses of
the heavy (right-handed (RH))
Majorana neutrinos. In the Higgs Triplet Model (HTM),
which is often called also ``type II see-saw model'',
the scale $\Lambda$ is related to the masses of the
new physical neutral, singly and doubly charged
Higgs particles.

  The scale $\Lambda$ at which the new physics, associated with the
existence of neutrino masses and mixing, manifests itself, in principle,
can have in the cases of type I seesaw and HT
models relevant for the present study an arbitrary large value
(see, e.g., \cite{seesaw,typeII,Raidal:2004vt}),
up to the GUT scale of $2\times 10^{16}$ GeV and even beyond,
up to the Planck scale. An interesting
possibility which can also be theoretically and
phenomenologically well motivated
both for the type I seesaw and HT models
is to have the new physics at the TeV scale, i.e.,
$\Lambda\sim (100 - 1000)$ GeV
(for the type I seesaw scenario see the discussions in, e.g.,
\cite{Shaposhnikov:2006nn,Gavela:2009cd,Ibarra:2011xn,Kersten:2007vk}).
In the TeV scale class of
type I see-saw models of interest,
the flavour structure of the couplings of the
new particles to the charged leptons is basically determined
by the requirement of reproducing the data on
the neutrino oscillation parameters
\cite{Raidal:2004vt,Shaposhnikov:2006nn,Gavela:2009cd,Ibarra:2011xn}.
In HTM these couplings are proportional
to the Majorana mass matrix of the left-handed flavour neutrinos.

As a consequence, the rates of
the LFV processes in the charged lepton sector can be calculated
in terms of a few parameters.
In the TeV scale type I seesaw scenario, for instance,
these parameters are constrained by different sets of data such as, e.g.,
data on neutrino oscillations, from EW precision tests and
on the LFV violating processes $\mu\rightarrow e+\gamma$,
$\mu\rightarrow 3e$, $\mu^- - e^-$ conversion
nuclei, etc. Nevertheless, the predicted rates
of the LFV charged lepton decays $\mu\rightarrow e+\gamma$,
$\mu^+\rightarrow e^+ e^- e^+$ and of the $\mu^- - e^-$ conversion
in both the TeV scale type I seesaw and HT models of interest
are within the reach of the future experiments searching
for lepton flavour violation
\footnote{Using the result given in eq. (\ref{muegDnus})
as a starting point,
it was shown in \cite{BPP77,ChengLi77,STPmu3e77}
that the rates of the LFV muon decays
$\mu\rightarrow e+\gamma$ and
$\mu\rightarrow 3e$, can be close to the existing upper limits
in theories with heavy neutral leptons (or heavy neutrinos, for that matter)
which have charged current weak interaction type couplings
to the electron and the muon. Although the specific model considered
in \cite{BPP77,ChengLi77,STPmu3e77} is not viable, the general
conclusion of these studies remains valid.
}
even when the parameters of the model do not allow
production of the new particles with observable rates
at the LHC~\cite{Ibarra:2011xn}.

 The role of the experiments searching for lepton flavour
violation to test and  constrain low scale see-saw models will be
significantly strengthened in the next years.
Searches for $\mu-e$ conversion at the planned COMET experiment
at KEK~\cite{comet} and Mu2e experiment
at Fermilab~\cite{mu2e} aim to reach sensitivity to
$\rm{CR}(\mu\, {\rm Al} \to e\, {\rm Al})\approx  10^{-16}$,
while, in the longer run, the PRISM/PRIME experiment in KEK~\cite{PRIME}
and the project-X experiment in Fermilab~\cite{projectX}
are being designed to probe values of the $\mu-e$ conversion rate
on ${\rm Ti}$, which are by 2 orders of magnitude smaller,
$\rm{CR}(\mu\, {\rm Ti} \to e\, {\rm Ti})\approx 10^{-18}$~\cite{PRIME}.
There are also plans to perform a new search for the
$\mu^+\rightarrow e^+ e^- e^+$ decay \cite{muto3eNext},
which will probe values of the corresponding
branching ratio down to
${\rm BR}(\mu^+\rightarrow e^+ e^- e^+) \approx 10^{-15}$,
i.e., by 3 orders of magnitude smaller than
the best current upper limit, eq.~(\ref{muto3eexp}).
Furthermore, searches for tau lepton flavour violation
at superB factories aim to reach a sensitivity to
${\rm BR}(\tau\rightarrow (\mu,e)\gamma)\approx 10^{-9}$,
while a next generation experiment on the
$\tau\to 3\mu$ decay is expected to reach sensitivity to
${\rm BR}(\tau\to 3\mu) = 10^{-10}$~\cite{Akeroyd:2004mj}.

 In the present article we investigate the LFV
$\tau$ decays $\tau\to (e,\mu)\gamma$ and $\tau\to 3\mu$
in the frameworks of the TeV scale type I see-saw and
HT models of neutrino mass generation.
We derive predictions for the rates of the indicated $\tau$ LFV decays
in the two models and analyse the possibilities of observation
of these decays in present and planned future experiments.

 Studies of the LFV $\tau$ decays
$\tau\rightarrow \mu \gamma$ and $\tau\rightarrow e \gamma$,
but not of the $\tau\to 3\mu$ decay, in the TeV scale
type I seesaw model where performed in \cite{Gavela:2009cd}.
In \cite{AAS_PRD79:2009} the authors investigated the
the $\tau\to 3\mu$ decay in the  Higgs Triplet model.
Comments about the $\tau\rightarrow \mu \gamma$,
$\tau\rightarrow e \gamma$ and $\tau\to 3\mu$ decays
in the Higgs Triplet model were made
in ref. \cite{KOS_PLB566:2003}.
However, our study of the $\tau$ LFV decays
overlaps little with those performed in
\cite{Gavela:2009cd,AAS_PRD79:2009,KOS_PLB566:2003}.

%
\section{TeV Scale Type I See-Saw Model}

%
\mathversion{bold}
\subsection{Brief Review of the Model}
\mathversion{normal}

 We denote the light and heavy Majorana neutrino
mass eigenstates of the
type I see-saw model \cite{seesaw}
as $\chi_i$ and $N_k$, respectively.
\footnote{We use the same notations as in
\cite{Ibarra:2011xn,Ibarra:2010xw}.}
The  charged and neutral current weak
interactions involving  the light and
heavy Majorana neutrinos have the form:
\begin{eqnarray}
\label{nuCC}
\mathcal{L}_{CC}^\nu
&=& -\,\frac{g}{\sqrt{2}}\,
\bar{\ell}\,\gamma_{\alpha}\,\nu_{\ell L}\,W^{\alpha}\;
+\; {\rm h.c.}
=\, -\,\frac{g}{\sqrt{2}}\,
\bar{\ell}\,\gamma_{\alpha}\,
\left( (1+\eta)U \right)_{\ell i}\,\chi_{i L}\,W^{\alpha}\;
+\; {\rm h.c.}\,,\\
\label{nuNC}
\mathcal{L}_{NC}^\nu &=& -\, \frac{g}{2 c_{w}}\,
\overline{\nu_{\ell L}}\,\gamma_{\alpha}\,
\nu_{\ell L}\, Z^{\alpha}\;
= -\,\frac{g}{2 c_{W}}\,
\overline{\chi_{i L}}\,\gamma_{\alpha}\,
\left (U^\dagger(1+2\eta)U\right)_{ij}\,\chi_{j L}\,
Z^{\alpha}\,,\\
\mathcal{L}_{CC}^N &=& -\,\frac{g}{2\sqrt{2}}\,
\bar{\ell}\,\gamma_{\alpha}\,(RV)_{\ell k}(1 - \gamma_5)\,N_{k}\,W^{\alpha}\;
+\; {\rm h.c.}\,
\label{NCC},\\
 \mathcal{L}_{NC}^N &=& -\frac{g}{4 c_{W}}\,
\overline{\nu_{\ell L}}\,\gamma_{\alpha}\,(RV)_{\ell k}\,(1 - \gamma_5)\,N_{k}\,Z^{\alpha}\;
+\; {\rm h.c.}\,.
\label{NNC}
\end{eqnarray}
%
Here $(1+\eta)U = U_{\rm PMNS}$ is the PMNS
neutrino mixing matrix \cite{BPont57,MNS62},
$U$ is a $3\times 3$ unitary
matrix which diagonalises the Majorana mass matrix
of the left-handed (LH) flavour neutrinos $\nu_{\ell L}$,
$m_{\nu}$, generated by the see-saw mechanism,
$V$ is the unitary matrix which diagonalises the
Majorana mass matrix of the heavy RH neutrinos and
the matrix $R$ is determined by (see \cite{Ibarra:2010xw})
$R^* \cong M_D\, M^{-1}_{N}$, $M_D$ and $M_N$ being the
neutrino Dirac and the RH neutrino Majorana mass matrices,
respectively, $|M_D| \ll |M_N|$.
The matrix $\eta$ characterises the deviations
from unitarity of the PMNS matrix:
\begin{equation}
	\eta\;\equiv\;-\half R R^\dagger =
-\half (RV)(RV)^\dagger = \eta^\dagger\,.
\label{eta}
\end{equation}
%
In the TeV scale type I see-saw model on interest,
the masses of the heavy Majorana neutrinos $N_k$, $M_k$,
are supposed to lie in the interval $M_k\sim (100 - 1000)$ GeV.
The couplings $(RV)_{lk}$ are bounded, in particular, by
their relation to the elements of the Majorana mass matrix
of LH flavour neutrinos $(m_{\nu})_{\ell\ell'}$, all of which have to
be smaller than approximately 1 eV:
\begin{equation}
|\sum_{k} (RV)^*_{\ell'k}\;M_k\, (RV)^{\dagger}_{k\ell}|
\cong |(m_{\nu})_{\ell'\ell}| \lesssim 1~{\rm eV}\,,
~\ell',\ell=e,\mu,\tau\,.
\label{VR1}
\end{equation}
%
These constraints can be satisfied
for sizeable values of the couplings
$|(RV)_{\ell k}|$ in a model with two heavy Majorana
neutrinos $N_{1,2}$, in which $N_1$ and $N_2$
have close masses forming a pseudo-Dirac state \cite{LWPD81,STPPD82},
$M_2 = M_1 (1 + z)$, $M_{1,2},z >0$, $z \ll 1$, and their couplings
satisfy \cite{Ibarra:2011xn}
\begin{equation}
(RV)_{\ell 2}=\pm i\, (RV)_{\ell 1}\sqrt{\frac{M_1}{M_2}}\,,~\ell=e,\mu,\tau\,.
\label{rel0}
\end{equation}
%
In this scenario with sizeable CC and NC couplings of $N_{1,2}$,
the requirement of reproducing the correct low
energy neutrino oscillation parameters
constrains significantly~\cite{Shaposhnikov:2006nn,Kersten:2007vk}
and in  certain cases determines the neutrino Yukawa couplings
\cite{Raidal:2004vt,Gavela:2009cd,Ibarra:2011xn}.
Correspondingly, the flavour dependence of the couplings
$(RV)_{\ell 1}$ and  $(RV)_{\ell 2}$ in  eqs. (\ref{NCC}) and  (\ref{NNC})
is also determined and in the case of interest takes
the form \cite{Ibarra:2011xn}:
\begin{eqnarray}
\label{mixing-vs-y}
\left|\left(RV\right)_{\ell 1} \right|^{2}&=&
\frac{1}{2}\frac{y^{2} v^{2}}{M_{1}^{2}}\frac{m_{3}}{m_{2}+m_{3}}
    \left|U_{\ell 3}+i\sqrt{m_{2}/m_{3}}U_{\ell 2} \right|^{2}\,,
~~{\rm NH}\,,\\
\left|\left(RV\right)_{\ell 1} \right|^{2}&=&
\frac{1}{2}\frac{y^{2} v^{2}}{M_{1}^{2}}\frac{m_{2}}{m_{1}+m_{2}}
    \left|U_{\ell 2}+i\sqrt{m_{1}/m_{2}}U_{\ell 1} \right|^{2}
\cong \;\frac{1}{4}\frac{y^{2} v^{2}}{M_{1}^{2}}
\left|U_{\ell 2}+iU_{\ell 1} \right|^{2}\,,
\,{\rm IH}\,,
\label{mixing-vs-yIH}
\end{eqnarray}
%
where $y$ represents the maximum eigenvalue of the neutrino
Yukawa matrix and $v\simeq174$ GeV.
In the last equation we have neglected the
$N_1 - N_2$ mass difference setting $z =0$
and used the fact that for the IH spectrum one has
$m_1 \cong m_2$. For $(RV)_{\ell 1,2}$ satisfying
eqs. (\ref{rel0}) - (\ref{mixing-vs-yIH}),
eq.~(\ref{VR1}) is automatically fulfilled.

 The low energy electroweak precision data
on processes involving light neutrinos
imply the following upper limits on the
couplings \cite{Antusch:2008tz,Akhmedov:2013hec}
(see also \cite{Antusch:2006}):
\begin{eqnarray}
 |(RV)_{e1}|^{2} & \lesssim & 2\times 10^{-3}
\label{e-bound}\,,\\
 |(RV)_{\mu 1}|^{2} &\lesssim & 0.8\times 10^{-3}
\label{mu-bound}\,,\\
|(RV)_{\tau 1}|^{2} & \lesssim & 2.6\times 10^{-3}
\label{tau-bound}\,.	
\end{eqnarray}
%

 Let us add finally that in the class of
type I see-saw models with
two heavy Majorana neutrinos we are considering
(see, e.g., \cite{3X2Models,Ibarra:2003up,PRST05}),
one of the three light (Majorana) neutrinos
is massless and hence the neutrino mass spectrum is
hierarchical (see, e.g., \cite{PDG2012}).
In the case of normal hierarchical (NH) spectrum we have
$m_{1}=0$, $m_{2}=\sqrt{\Delta m^2_{21}}$ and $m_{3}=\sqrt{\Delta m^2_{31}}$,
while if the spectrum is inverted hierarchical (IH),
$m_{3}=0$, $m_{2}=\sqrt{|\Delta m^2_{32}|}$ and
$m_{1}=\sqrt{|\Delta m^2_{32}|-\Delta m^2_{21}}
\cong \sqrt{|\Delta m^2_{32}|}$,
with $\Delta m^2_{32} =  m^2_{3} -  m^2_{2} < 0$.
In both cases we have: $\Delta m^2_{21}/\Delta m^2_{31(23)} \cong 0.03 \ll 1$.

%
\mathversion{bold}
\subsection{The $\tau\rightarrow \mu\gamma$ and
$\tau\rightarrow e\gamma$ Decays}
\mathversion{normal}
 In the type I see-saw scheme of interest with two heavy Majorana neutrinos,
the ratio of the decay rates $\Gamma(l_\alpha\rightarrow l_\beta\gamma)$ and
$\Gamma(l_\alpha\rightarrow \nu_\alpha l_\beta\overline{\nu}_\beta)$
can be written as \cite{BPP77,ChengLi77,Ibarra:2011xn}:
\begin{equation}
\frac{\Gamma(l_\alpha\rightarrow l_\beta\gamma)}
{\Gamma(l_\alpha\rightarrow \nu_\alpha l_\beta\overline{\nu}_\beta)}
=\frac{3\alpha_{\rm em}}{32\pi}|T|^2\,,
\label{FracGG}
\end{equation}
%
where
%
\begin{eqnarray}
T\approx2|(RV)^*_{\beta 1}(RV)_{\alpha 1}|\,|G(x)-G(0)|\,,\\
\label{Tdef}
G(x)=\frac{10-43 x+78 x^2-49 x^3+4 x^4+18 x^3\ln{x}}{3(x-1)^4}\,,
\label{FracGG2}
\end{eqnarray}
%
with $x=M_1^2/M_W^2$. In deriving eq. (\ref{Tdef}) we have used
the relation (\ref{rel0}) and have neglected the $N_1 - N_2$
mass difference. The $l_\alpha\rightarrow l_\beta\gamma$ decay branching
ratio is given by:
\begin{equation}
{\rm BR}(l_\alpha\rightarrow l_\beta\gamma)=\frac{\Gamma(l_\alpha\rightarrow l_\beta\gamma)}{\Gamma(l_\alpha\rightarrow \nu_\alpha l_\beta\overline{\nu}_\beta)}{\rm Br}(l_\alpha\rightarrow \nu_\alpha l_\beta\overline{\nu}_\beta),
\label{FracGG}
\end{equation}
%
with ${\rm BR}(\mu\rightarrow\nu_{\mu}\,e~\overline{\nu}_e)\approx 1$,
${\rm BR}(\tau\rightarrow\nu_{\tau}\,
\mu~\overline{\nu}_{\mu})=0.1739$, and
${\rm BR}(\tau\rightarrow\nu_{\tau}\,e~\overline{\nu}_{e})= 0.1782$
\cite{PDG2012}.

 The predictions of the model under discussion for
${\rm BR}(\mu\to e\gamma)$ and the
constraints on the product of couplings $|(RV)^*_{e 1}(RV)_{\mu 1}|$,
as well as on the Yukawa coupling $y$, following
from the experimental upper limit on  ${\rm BR}(\mu\to e\gamma)$,
were discussed in detail in \cite{Ibarra:2011xn,Dinh:2012bp}.
Here we concentrate on the phenomenology of the $\tau\to \mu\gamma$
and $\tau\to e\gamma$ decays. Using the current upper limits on
${\rm BR}(\tau\to \mu\gamma)$
and ${\rm Br}(\tau\to e\gamma)$
quoted in eqs. (\ref{tautomugexp}) and (\ref{tautoegexp}),
we obtain the following
upper bounds:
\begin{eqnarray}
\label{BabarLimit}
&&\tau\to\mu\gamma:~~~|(RV)^*_{\mu 1}(RV)_{\tau 1}|\leq 2.7\times 10^{-2}~(0.9\times 10^{-2})~~~M_1=100~(1000)\,{\rm GeV}\,,\\
&&\tau\to e\gamma:~~~~|(RV)^*_{e 1}(RV)_{\tau 1}|\leq2.3\times 10^{-2}~(0.8\times 10^{-2})~~~M_1=100~(1000)\,{\rm GeV}.
\end{eqnarray}
%
These constraints are weaker than those implied by the limits
quoted in eqs. (\ref{e-bound}) - (\ref{tau-bound}).
The planned experiments at the SuperB factory, which are expected to
probe values of ${\rm BR}(\tau\to (\mu,e)\gamma)\geq 10^{-9}$,
will be sensitive to
\begin{eqnarray}
\label{superB}
\tau\to(\mu,e)\gamma:~~~|(RV)^*_{(\mu,e) 1}(RV)_{\tau 1}|\geq4.0\times 10^{-3}~(1.4\times 10^{-3})~~M_1=100~(1000)\,{\rm GeV}\,.
\end{eqnarray}
%
The minimal values quoted above are of the same order as the upper limits following from
the constraints  (\ref{e-bound}) - (\ref{tau-bound}).

  The $\tau$ decay branching ratios of interest depend on the neutrino mixing
parameters via the quantity $|(RV)*_{l 1}(RV)_{\tau 1}|$, $l=e,\mu$.
In the  case of NH neutrino mass spectrum,
$|(RV)_{l 1}|\propto |U_{l 3} + i\sqrt{m_2/m_3}\,U_{l 2}|$
is different from zero for any values of
the neutrino mixing parameters from their
$3\sigma$ experimentally determined allowed ranges and
for any $l=e,\,\mu,\,\tau$.
This implies that there cannot be further suppression of the
$\tau\to(\mu,e)\gamma$ decay rates due to a cancellation between
the terms in the expressions for $|(RV)_{l 1}|$.

  In contrast, depending on the values of the Dirac
and Majorana CPV phases $\delta$ and $\alpha_{21}$
of the PMNS matrix, we can have strong suppression of
the couplings
$|(RV)_{l 1}|$, $l=e,\mu$,
which enter into the expressions for
${\rm BR}(\tau\to (\mu,e)\gamma)$
if the neutrino mass spectrum is of the IH type
\cite{Ibarra:2011xn,Dinh:2012bp}.
Indeed, in this case we have
$|(RV)_{l 1}|\propto |U_{l 3} + i\,U_{l 2}|$, $l=e,\mu,\tau$.
For $\alpha_{21} = -\,\pi$,  $|U_{e 3} + i\,U_{e 2}|$ can be rather small:
$| U_{e 2} + i U_{e 1}|^{2} = c^2_{13}(1 - \sin2\theta_{12})\cong 0.0765$,
where we have used the best fit values of $\sin^2\theta_{12} = 0.307$ and
$\sin^2\theta_{13} = 0.0236$. As was shown in \cite{Dinh:2012bp},
we can have $| U_{\mu 2} + i U_{\mu 1}|^{2} = 0$ for specific
values of $\delta$ lying the interval $0\leq \delta \lesssim 0.7$.
In this case the value of the phases $\alpha_{21}$ is determined
by the values of $\delta$ and $\theta_{12}$ (for further details see
\cite{Dinh:2012bp}).

  We analyse next the possibility of having strongly
suppressed coupling $|(RV)_{\tau 1}|^2$, i.e., to have
$|(RV)_{\tau 1}|^2\propto |U_{\tau 2} + i U_{\tau 1}|^{2} = 0$,
in the case of IH spectrum. The suppression in question
can take place if
\begin{equation}
\sin\theta_{13} = \frac{s_{12} - c_{12}\, \sin \frac{\alpha_{21}}{2}}
{c_{12}\, \cos\delta  + s_{12}\, \sin(\delta + \frac{\alpha_{21}}{2})}\,\tan\theta_{23}\,,
\label{th13RVtau10}
\end{equation}
%
and if in addition the values of the phases $\delta$ and $\alpha_{21}$
are related via the equation:
\begin{equation}
c_{12}\,s_{23}\,\cos\frac{\alpha_{21}}{2} -
c_{23}\,s_{13}\left [ c_{12}\,\sin\delta
- s_{12}\,\cos(\delta + \frac{\alpha_{21}}{2} \right ] = 0\,.
\label{RVtau10}
\end{equation}
%
One simple solution to eq. (\ref{RVtau10}) obviously
is $\delta = \alpha_{21} =\pi$. For these values of
 $\delta$ and $\alpha_{21}$, eq. (\ref{th13RVtau10})
becomes:
\begin{equation}
\sin\theta_{13} = \frac{c_{12} - s_{12}}
{c_{12}\, + s_{12}}\,\tan\theta_{23}\,.
\label{th13RVtau101}
\end{equation}
%
Using the the best fit values of $\sin^2\theta_{12}$ and
$\sin^2\theta_{23}$ quoted in eq. (\ref{thetaosc}),
we get from eq. (\ref{th13RVtau101}):
$\sin\theta_{13} = 0.162$, which is very close
to the best fit value of 0.155~(0.156)
quoted in eq. (\ref{thetaosc}).
For $|U_{\tau 2} + i U_{\tau 1}|^{2} \cong 0$,
all LFV decays of the $\tau$ charged lepton,
including  $\tau^-\rightarrow \mu^- + \mu^+ + \mu^-$,
$\tau^- \rightarrow \mu^- + e^+ + e^-$, etc.,
in the TeV scale type I seesaw  model
we are considering will be strongly suppressed.
%
\mathversion{bold}
\subsection{The $\tau\rightarrow 3\mu$ Decay}
\mathversion{normal}
 We have obtained the $\tau\rightarrow 3\mu$ decay rate by
adapting the result of the calculation of the $\mu\to 3e$
decay rate performed in  \cite{IP_NuclB:1995} in a scheme with
heavy RH neutrinos and type I seesaw mechanism of neutrino mass generation.
After recalculating the form factors and neglecting the
corrections $\sim m_\mu/m_\tau \cong 0.06$
and the effects of the difference between the masses of
$N_1$ and $N_2$,
we find in the model of interest to leading order in the
small parameters $|(RV)_{l 1}|$:
\begin{eqnarray}
\label{t3mu}
{\rm BR}(\tau \to 3\mu)
&=& \frac{\alpha_{em}^2}{16\pi^2\sin^4{\theta_W}}\left| (RV)_{\tau1}^{*} (RV)_{\mu1}\right|^2\,
\left |C_{\tau3\mu}(x)\right |^2\times {\rm BR}(\tau\to\mu\bar{\nu}_\mu\nu_\tau),\\
\nonumber
\left |C_{\tau3\mu}(x)\right |^2 &=& 2\left|\frac{1}{2}F_B^{\tau3\mu}+F_z^{\tau3\mu}-2\sin^2{\theta_W}(F_z^{\tau3\mu}-F_{\gamma})\right|^2
+4\sin^4{\theta_W}\left|F_z^{\tau3\mu}-F_{\gamma}\right|^2\\
\nonumber
&&+16\sin^2{\theta_W}
\left[(F_z^{\tau3\mu}+\frac{1}{2}F_B^{\tau3\mu})G_{\gamma}\right]
- 48\sin^4{\theta_W}\left[(F_z^{\tau3\mu}-F_\gamma^{\tau3\mu})G_{\gamma}\right]\\
&&+32\sin^4{\theta_W}|G_\gamma|^2\left(\log{\frac{m_\tau^2}{m_\mu^2}}-\frac{11}{4}\right)\,.
\label{cm3t}
\end{eqnarray}
%
Here
\begin{eqnarray}
&&F_z^{\tau3\mu}(x)=F_z(x)+2G_z(0,x),\,\,\,F_B^{\tau3\mu}(x)=-2(F_{XBox}(0,x)-F_{XBox}(0,0)),\\
&&F_{\gamma}(x)=\frac{x(7x^2-x-12)}{12 (1-x)^3}-\frac{x^2(12-10x+x^2)}{6(1-x)^4}\log{x},\\
&&G_{\gamma}(x)=-\frac{x(2x^2+5x-1)}{4 (1-x)^3}-\frac{3x^3}{2(1-x)^4}\log{x},\\
&&F_z(x)=-\frac{5x}{2(1-x)}-\frac{5x^2}{2(1-x)^2}\log{x},\\
&&G_z(x,y)=-\frac{1}{2(x-y)}\left[\frac{x^2(1-y)}{(1-x)}\log{x}-\frac{y^2(1-x)}{(1-y)}\log{y}\right],
\end{eqnarray}
\begin{eqnarray}
\nonumber
&&F_{XBox}(x,y)=-\frac{1}{x-y}\left\{(1+\frac{x y}{4})\left[\frac{1}{1-x}+\frac{x^2}{(1-x)^2}\log{x}-\frac{1}{1-y}-\frac{y^2}{(1-y)^2}\log{y}\right]\right.\\ &&~~~~\left.-2xy\left[\frac{1}{1-x}+\frac{x}{(1-x)^2}\log{x}-\frac{1}{1-y}-\frac{y}{(1-y)^2}\log{y}\right]\right\}.
\end{eqnarray}
%
In writing the expression for ${\rm BR}(\tau \to 3\mu)$ in eq. (\ref{t3mu})
we have used for the decay rate $\Gamma(\tau\to\mu\bar{\nu}_\mu\nu_\tau) =
G^2_F m^5_{\tau}/(192\pi^3)$.

 The factor  $|C_{\tau3\mu}(x)|^2$ in the expression for
${\rm BR}(\mu \to 3\mu)$  is a monotonically increasing function
of the heavy Majorana neutrino mass $M_1$.
The dependence of  $|C_{\tau3\mu}(x)|^2$ on $M_1$ is shown in Fig. 1.
At  $M_1 = 100~(1000)$ GeV, the function  $|C_{\tau3\mu}(x)|^2$
has values 1.53 (36.85).

  The present experimental limit on ${\rm BR}(\tau\to 3\mu)$,
eq. (\ref{tauto3muexp}), leads to a weaker constraint than that
following from the upper limits quoted in
eqs. (\ref{mu-bound}) and (\ref{tau-bound}):
\begin{equation}
|(RV)_{\tau1}^{*} (RV)_{\mu1}|<1.1\times 10^{-1}~
(2.3\times 10^{-2})~{\rm for}~{\rm M_1=100~(1000)~GeV}.
\end{equation}
%
The next generation of experiments will be sensitive to
${\rm BR}(\tau\to 3\mu) \geq 10^{-10}$, and thus to:
\begin{equation}
\label{LimitRV}
|(RV)_{\tau1}^{*} (RV)_{\mu1}|\geq 7.7\times 10^{-3}~
(1.6\times 10^{-3})~{\rm for}~{\rm M_1=100~(1000)~GeV}\,.
\end{equation}
%
As we see, in the case of $M_1 = 1000$ GeV, the minimal value of
$|(RV)_{\tau1}^{*} (RV)_{\mu1}|$ to which the future
planned experiments will be sensitive
is of the order of the upper bound on
$|(RV)_{\tau1}^{*} (RV)_{\mu1}|$ following
from the limits (\ref{mu-bound}) and (\ref{tau-bound}).

 Consider next the dependence of the decay rate on the CPV phases and the
neutrino oscillation parameters. In the case of NH mass spectrum we have:
\begin{equation}
{\rm BR}(\tau\to 3\mu)\propto |(RV)_{\tau1}^{*} (RV)_{\mu1}|^2\propto
|U_{\tau 3}+i\sqrt{\frac{m_2}{m_3}}U_{\tau 2}|^2\,
|U_{\mu 3}+i\sqrt{\frac{m_2}{m_3}}U_{\mu 2}|^2\,.
\end{equation}
%
Using the best fit values of the neutrino mixing angles
and mass squared differences,
quoted in eqs. (\ref{Deltam2osc}) and  (\ref{thetaosc})
and varying the Dirac and Majorana CPV phases in
the interval of $[0,2\pi]$, we find that
$|U_{\mu 3}+i\sqrt{m_2/m_3}U_{\mu 2}||U_{\tau 3}+i\sqrt{m_2/m_3}U_{\tau 2}|$
takes values in the interval $(0.31 - 0.59)$.
It follows from this result and the inequality (\ref{LimitRV}) that
the future experiments on the  $\tau\to 3\mu$ decay will be
sensitive to values of the Yukawa coupling
$y \geq 0.10~(0.46)~{\rm for}~M_1=100~(1000)$ GeV.
The minimal values in these lower limits are larger
than the upper limits on $y$ following from the current
upper bound (\ref{mutoegexp}) on
${\rm BR}(\mu \to e +\gamma)$
\cite{Dinh:2012bp}.

  A suppression of the $\tau\to 3\mu$ decay rate might
occur in the case of IH mass
due to possible cancellations between the terms in the factors
$|(RV)_{\mu1}|$ and  $|(RV)_{\tau1}|$, as was discussed in
the previous subsection. Using again the best fit values
of the neutrino oscillation parameters
and varying the leptonic CPV phases in the interval $[0,2\pi]$,
we find $0.003 \leq |U_{\mu 2}+iU_{\mu 1}||U_{\tau 2}+iU_{\tau 1}| \leq 0.51$.
Thus, in the case of IH spectrum, the future experiments with sensitivity to
${\rm BR}(\tau\to 3\mu) \geq 10^{-10}$ will probe values of
$y \geq 0.14~(0.64)~{\rm for}~M_1=100~(1000)$ GeV.
Again the minimal values in these lower limits are larger
than the upper limits on
$y$ following from the current upper bound (\ref{mutoegexp}) on
${\rm BR}(\mu \to e +\gamma)$ \cite{Dinh:2012bp}. \\

  For specific values of, e.g., the CPV phases of the neutrino mixing matrix
one can obtain more stringent upper bounds than those already
discussed on the branching ratios of the $\tau\to \mu +\gamma$,
$\tau \to  e + \gamma$ and  $\tau\to 3\mu$ decays
due to their relation to the  $\mu\to e +\gamma$
decay branching ratio and the fact that the latter is
severely constrained. Indeed, it follows from
eqs. (\ref{FracGG}), (\ref{FracGG2}) and
(\ref{t3mu}) that we have:
\begin{eqnarray}
\label{Rtegmueg}
\frac{{\rm BR}(\tau\to e +\gamma)}
{{\rm BR}(\mu\to e+\gamma)} = \frac{|(RV)_{\tau 1}|^2}{|(RV)_{\mu 1}|^2}\,
BR(\tau\to e\bar{\nu}_e \nu_{\tau})\,,\\
\label{Rtmugmueg}
\frac{{\rm BR}(\tau\to \mu +\gamma)}
{{\rm BR}(\mu\to e+\gamma)} = \frac{|(RV)_{\tau 1}|^2}{|(RV)_{e 1}|^2}\,
BR(\tau\to \mu \bar{\nu}_\mu \nu_{\tau})\,,\\
\label{Rt3mumu3e}
\frac{{\rm BR}(\tau\to 3\mu)}
{{\rm BR}(\mu\to 3e)} = \frac{|(RV)_{\tau 1}|^2}{|(RV)_{e 1}|^2}\,
BR(\tau\to \mu \bar{\nu}_\mu \nu_{\tau})\ =
\frac{{\rm BR}(\tau\to \mu +\gamma)}
{{\rm BR}(\mu\to e+\gamma)}\,,\\
\label{Rt3mumueg}
\frac{{\rm BR}(\tau\to 3\mu)}
{{\rm BR}(\mu\to e + \gamma)} =  \frac{\alpha_{em}}{6\pi \sin^4\theta_W}
\frac{|C_{\tau3\mu}(x)|^2}{|G(x)-G(0)|^2}\,
\frac{|(RV)_{\tau 1}|^2}{|(RV)_{e 1}|^2}\, BR(\tau\to \mu \bar{\nu}_\mu \nu_{\tau})\,.
\end{eqnarray}
%
The explicit expressions for $|(RV)_{l 1}|^2$,
eqs. (\ref{mixing-vs-y}) and (\ref{mixing-vs-yIH}),
imply that the ratios of interest
in eqs. (\ref{Rtegmueg}) - (\ref{Rt3mumu3e}) do not depend on
the heavy Majorana neutrino mass $M_1$ and on the Yukawa coupling $y$
and are determined by the values of the neutrino oscillation parameters
and of the CPV phases in the neutrino mixing matrix,
as was noticed also in ref. \cite{Gavela:2009cd}.
Using the best fit values quoted in eqs. (\ref{Deltam2osc}) and
(\ref{thetaosc}) and varying the Dirac and Majorana phases
in the interval $[0,2\pi]$ we obtain in the case of
NH neutrino mass spectrum:
\begin{eqnarray}
\label{RtegmuegNH}
0.37 \leq \frac{|(RV)_{\tau 1}|^2}{|(RV)_{\mu 1}|^2} \leq 9.06 \,,\\
\label{RtmugmuegNH}
1.90 \leq  \frac{|(RV)_{\tau 1}|^2}{|(RV)_{e 1}|^2} \leq 191.82\,,
\end{eqnarray}
%
In a similar way, we get in the case of IH neutrino mass spectrum:
\begin{eqnarray}
\label{RtegmuegIH}
4.84\times 10^{-4} \leq \frac{|(RV)_{\tau 1}|^2}{|(RV)_{\mu 1}|^2} \leq 15.13 \,,\\
\label{RtmugmuegIH}
3.25\times 10^{-4} \leq  \frac{|(RV)_{\tau 1}|^2}{|(RV)_{e 1}|^2} \leq 0.56\,.
\end{eqnarray}
%
Thus, in the case of the best fit values of the
neutrino oscillation parameters we always have
\begin{eqnarray}
\label{maxRtegmuegIH1}
{\rm BR}(\tau\to e +\gamma) \lesssim 2.67\times {\rm BR}(\mu\to e+\gamma)
< 1.52\times 10^{-12}\,,\\
\label{maxRtmugmuegNH1}
{\rm BR}(\tau\to \mu +\gamma) \lesssim 33.36\times {\rm BR}(\mu\to e+\gamma)
< 1.90\times 10^{-11}\,,
\end{eqnarray}
%
where we have used the current upper bound on
${\rm BR}(\mu\to e+\gamma)$, eq. (\ref{mutoegexp}).
The limits in eqs. (\ref{maxRtegmuegIH1}) and
(\ref{maxRtmugmuegNH1}) correspond respectively to the
IH and NH spectra. These values are beyond the expected
sensitivity reach of the planned future experiments.

 Using the $2\sigma$ ($3\sigma$) allowed ranges of the
neutrino oscillations parameters in the case of NH
neutrino mass spectrum
we obtain larger intervals of allowed values of
the ratios of interest:
\begin{eqnarray}
\label{RtegmuegNH2s}
NH:~~0.26~(0.08)\leq \frac{|(RV)_{\tau 1}|^2}{|(RV)_{\mu 1}|^2}
\leq 14.06~(16.73) \,,\\
\label{RtmugmuegNH2s}
NH:~~1.39~(0.53) \leq  \frac{|(RV)_{\tau 1}|^2}{|(RV)_{e 1}|^2}
\leq 497.74~(980.32)\,.
\end{eqnarray}
%
The maximal value of $|(RV)_{\tau 1}|^2/|(RV)_{e 1}|^2$
correspond to $\sin^2\theta_{12} = 0.275~(0.259)$,
$\sin^2\theta_{23} = 0.359~(0.348)$,
$\sin^2\theta_{13} = 0.0298~(0.0312)$,
$\delta = 0.203~(0.234)$, $\alpha_{21} = 6.199~(3.560)$ and
$\alpha_{31} = 3.420~(0.919)$.
At these values of the neutrino mixing parameters we have
$|(RV)_{\mu 1}|^2|(RV)_{e 1}|^2
\cong 6.98\times10^{-4}~(3.41\times10^{-4})\,y^4v^4/(16M^4_1)$,
$|(RV)_{\tau 1}|^2|(RV)_{\mu 1}|^2\cong 0.347~(0.335)\,y^4v^4/(16M^4_1)$.
Thus, the bound on  ${\rm BR}(\mu\to e +\gamma)$,
eq. (\ref{mutoegexp}), is satisfied for $M_1 = 100$ GeV
if $y^4v^4/(16M^4_1)\lesssim 2.29~(4.69)\times 10^{-6}$,
and for $M_1 = 1000$ GeV provided
$y^4v^4/(16M^4_1)\lesssim 2.68~(5.48)\times 10^{-7}$.
This implies that $|(RV)_{\tau 1}|^2|(RV)_{\mu 1}|^2 \lesssim 7.95~(15.7)\times 10^{-7}$ if $M_1 = 100$ GeV,
and $|(RV)_{\tau 1}|^2|(RV)_{\mu 1}|^2 \lesssim 9.30~(18.4)\times 10^{-8}$
for $M_1 = 1000$ GeV. The bound for $M_1 = 1000$ GeV is a
stronger constraint than that following from
the limits (\ref{mu-bound}) and  (\ref{tau-bound}).

  Using the inequalities in eqs. (\ref{RtegmuegNH2s}) and
(\ref{RtmugmuegNH2s}) we obtain:
\begin{eqnarray}
\label{maxRtegmuegNH2s2}
{\rm BR}(\tau\to e +\gamma) \lesssim
2.50~(2.98) \times {\rm BR}(\mu\to e+\gamma) < 1.43~(1.70) \times 10^{-12}\,,\\
\label{maxRtmugmuegNH2s2}
{\rm BR}(\tau\to \mu +\gamma) \lesssim
86.56~(170.48)\times {\rm BR}(\mu\to e+\gamma) < 4.93~(9.72)\times 10^{-11}\,.
\end{eqnarray}
%
These are the maximal values of ${\rm BR}(\tau\to e +\gamma)$
and ${\rm BR}(\tau\to \mu +\gamma)$, allowed
by the current upper bound on the $\mu\to e+\gamma$
decay rate in the TeV scale type I seesaw model considered
and in the case of NH neutrino mass spectrum.
If the $\tau\to e +\gamma$ and/or $\tau\to \mu +\gamma$ decays
are observed to proceed with branching ratios which are larger than
the bounds quoted above and it is established that the neutrino
mass spectrum is of the NH type, the model under discussion
will be strongly disfavored, if not ruled out.

 Performing a similar analysis in the case of IH spectrum
by employing the $2\sigma$ ($3\sigma$) allowed ranges of
the neutrino oscillations parameters we get:
\begin{eqnarray}
\label{RtegmuegIH2s}
IH:~~0.0~(0.0) \leq \frac{|(RV)_{\tau 1}|^2}{|(RV)_{\mu 1}|^2} < \infty~(\infty)  \,,\\
\label{RtmugmuegIH2s}
IH:~~0.0~(0.0) \leq  \frac{|(RV)_{\tau 1}|^2}{|(RV)_{e 1}|^2} \leq 0.64~(0.83)\,.
\end{eqnarray}
%
The infinity in eq. (\ref{RtegmuegIH2s}) corresponds to
$|(RV)_{\mu 1}| = 0$,
$|(RV)_{\tau 1}|\neq 0$, i.e., to very strongly suppressed
${\rm BR}(\mu\to e+\gamma)$ and ${\rm BR}(\tau\to \mu +\gamma)$.
One obtains $|(RV)_{\mu 1}| = 0$ for the following
values of the neutrino mixing angles from the
$2\sigma$ allowed intervals, and of the CPV phases:
$\sin^2\theta_{12} = 0.340$,
$\sin^2\theta_{23} = 0.547$,
$\sin^2\theta_{13} = 0.0239$,
$\delta = 6.185$, $\alpha_{21} = 3.077$ and
$\alpha_{31} = 4.184$ (i.e.,
$\delta \cong 2\pi$, $\alpha_{21} \cong \pi$ and
$\alpha_{31} \cong 1.3\pi$).
For  $|(RV)_{\mu 1}| = 0$,
the branching ratios  ${\rm BR}(\tau\to e +\gamma)$ and
${\rm BR}(\mu\to e+\gamma)$ are ``decoupled''.
Correspondingly, the upper bound on
${\rm BR}(\tau\to e +\gamma)$ is determined
in this case by the limits quoted in
eqs. (\ref{e-bound}) and  (\ref{tau-bound})
and has already been discussed by us.

 Using the same strategy and eq. (\ref{Rt3mumu3e}),
we obtain the constraint on ${\rm BR}(\tau\to 3\mu)$ following
from  the upper bound on ${\rm BR}(\mu\to 3e)$ at the best fit values,
$2\sigma$ ($3\sigma$) allowed ranges of the neutrino oscillation parameters:
\begin{eqnarray}
\label{Relate3mu3eBF}
{\rm BR}(\tau\to 3\mu) \lesssim
33.36\times {\rm BR}(\mu\to 3e) < 3.34\times 10^{-11}\,,\\
\label{Relate3mu3e23sig}
{\rm BR}(\tau\to 3\mu) \lesssim
86.56~(170.48)\times {\rm BR}(\mu\to 3e) < 8.66~(17.0)\times 10^{-11}\,.
\end{eqnarray}
%

The relation between ${\rm BR}(\tau\to 3\mu)$ and
${\rm BR}(\mu\to e\gamma)$ is somewhat less straightforward,
since it involves the $M_1$ dependent factor $C_0(x)$:
\begin{equation}
C_0(x)=\frac{\alpha_{em}}{6\pi \sin^4\theta_W}
\frac{|C_{\tau3\mu}(x)|^2}{|G(x)-G(0)|^2}.
\end{equation}
%
\begin{figure}[t]
\begin{center}
\includegraphics[width=16cm,height=10cm]{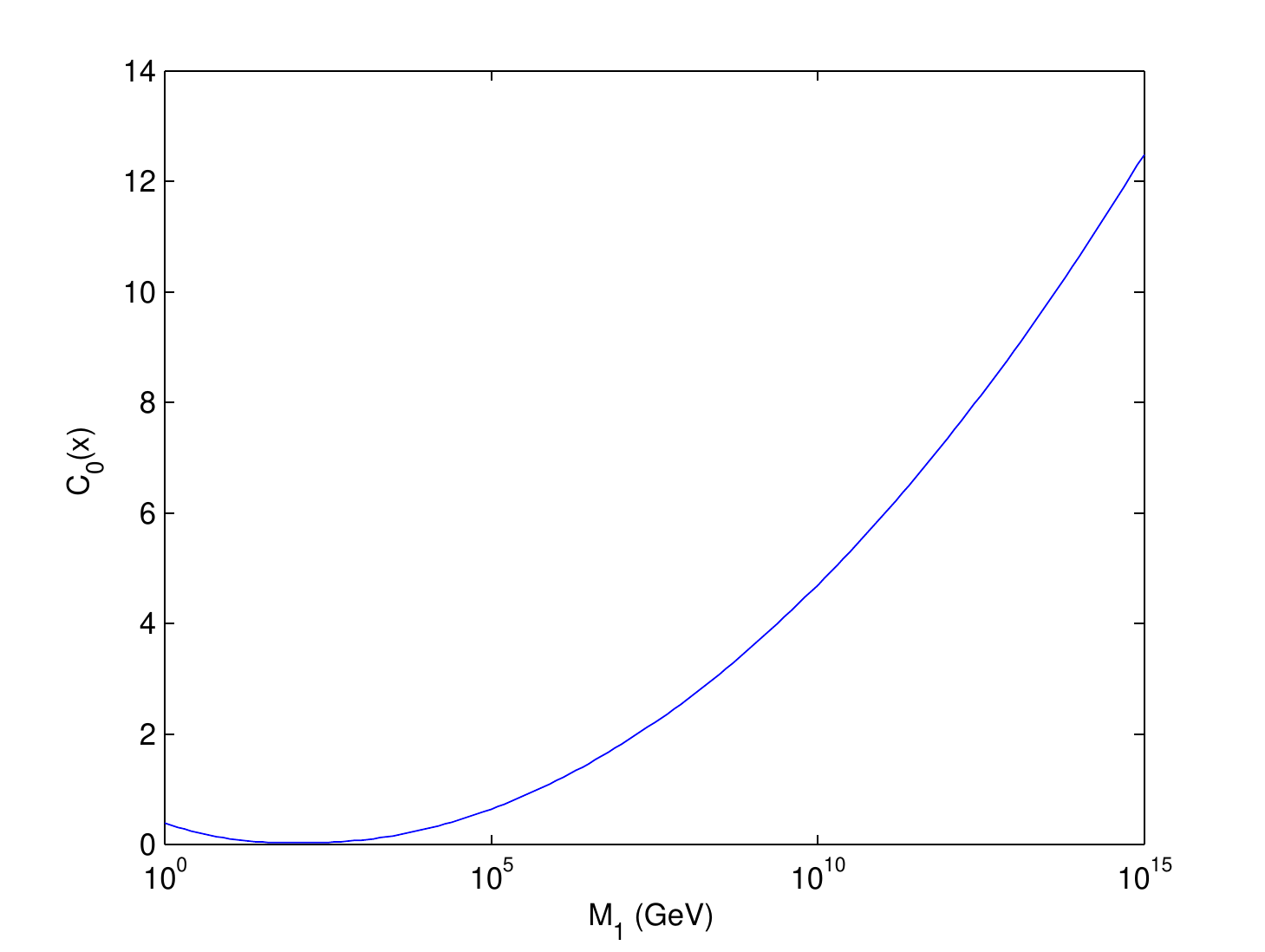}
\caption{The dependence of $C_{0}(x)$  as a function of the
see-saw mass scale $M_1$.
}
\label{C0x}
\end{center}
\end{figure}
For $50~{\rm GeV}\leq M_1\leq 1000~{\rm GeV}$,
$C_0(x)$ has its maximum of 0.0764 at $M_1=1000$ GeV.
This leads to
\begin{eqnarray}
\label{Relate3muegammaBF}
{\rm BR}(\tau\to 3\mu) \lesssim
2.55\times {\rm BR}(\mu\to e+\gamma) < 1.45\times 10^{-12}\,,\\
\label{Relate3muegamma23sig}
{\rm BR}(\tau\to 3\mu) \lesssim
6.61~(13.02)\times {\rm BR}(\mu\to e+\gamma) < 3.77~(7.42)\times 10^{-12}\,.
\end{eqnarray}
%
Thus, for $M_1$ having a value in the interval [50, 1000] GeV,
the branching ratio ${\rm BR}(\tau\to 3\mu)$
is predicted to be beyond the sensitivity reach of
$\sim 10^{-10}$ of the planned next generation experiment.
The observation of the $\tau\to 3\mu$ decay with
a branching ratio ${\rm BR}(\tau\to 3\mu)$
which is definitely larger than the
upper bounds quoted in eq. (\ref{Relate3muegamma23sig})
would strongly disfavor (if not rule out)
the TeV scale type I seesaw model under discussion
with $M_1 \sim (50 - 1000)$ GeV.

 It should be added that for $M_1 \geq 10^{3}$ GeV, the factor
$C_0(x)$ is a monotonically (slowly) increasing function of
$M_1$ (see Fig. \ref{C0x}). The upper bound  on
${\rm BR}(\tau\to 3\mu)$ following from the upper
bound on ${\rm BR}(\mu\to e+\gamma)$
and the $3\sigma$ ranges of the neutrino oscillation parameters,
can be bigger than $10^{-10}$ if $C_0(x) \geq 1.8$, which
requires $M_1 \geq 8.5\times 10^{6}$ GeV.  However,
the rates of the processes of interest
scale as  $\propto (v/M_1)^4$ and
at values of   $M_1 \geq 8.5\times 10^{6}$ GeV
are too small to be observed in the currently
planned experiments.

%
\section{The TeV Scale Higgs Triplet (Type II See-Saw) Model}
%
%
\mathversion{bold}
\subsection{Brief review of the TeV Scale Higgs Triplet Model}
\mathversion{normal}

 In its simplest version the Higgs Triplet Model (HTM)
\cite{typeII} is an extension of the SM, which contains one
additional $SU(2)_L$ triplet scalar field
$\Delta$ carrying two units of the weak
hypercharge $Y_W$. The Lagrangian of the HTM has
the form
~\footnote{We do not give here, for simplicity,
all the quadratic and quartic
terms present in the scalar potential
(see, $e.g.$, \cite{Akeroyd:2009nu}).
}:
\begin{eqnarray}
	\mathcal{L}^{\rm II}_{\rm seesaw} &=& -M_{\Delta}^{2}\,{\rm Tr}\left(\Delta^{\dagger}\Delta\right)
		-\left(h_{\ell\ell^{\prime}}\,\overline{\psi^{C}}_{\ell L}\,i\tau_{2}\,\Delta\,\psi_{\ell^{\prime}L}
		\,+\,\mu_{\Delta}\, H^{T}\,i\tau_{2}\,\Delta^{\dagger}\,H\,+\,{\rm h.c.}\right)\,,
\label{LtypeII}
\end{eqnarray}
%
where $(\psi_{\ell L})^T \equiv (\nu^T_{\ell L}~~\ell^T_{L})$,
$\overline{\psi^{C}}_{\ell L}
\equiv ( -\,\nu^T_{\ell L}C^{-1}~~-\,\ell^T_{L}C^{-1})$,
and $H$ are, respectively, the SM lepton and Higgs doublets,
$C$ being the charge conjugation matrix, and
\begin{equation}
	\Delta\;=\;\left(
			\begin{array}{cc}
				\Delta^{+}/\sqrt{2} & \Delta^{++} \\
				\Delta^{0}	& -\Delta^{+}/\sqrt{2}
			\end{array}\right)\,.
\end{equation}
%
In eq. (\ref{LtypeII}), $\mu_{\Delta}$ is a real parameter
characterising the soft explicit breaking of the total lepton charge
conservation. We will consider the TeV scale version of  HTM,
where the new physics scale $M_{\Delta}$,
associated with the mass of $\Delta$, takes values
$100~{\rm GeV}\lesssim M_{\Delta} \lesssim 1~{\rm TeV}$,
which, in principle, can be probed by LHC~\cite{colliders}.

   The light neutrino mass matrix $m_{\nu}$
is generated when the neutral component of $\Delta$
develops a ``small'' vev  $v_{\Delta} \propto \mu_{\Delta}$:
\begin{equation}
\left(m_{\nu}\right)_{\ell\ell^{\prime}}\, \equiv m_{\ell\ell^{\prime}}\,
\simeq\;2\,h_{\ell\ell^{\prime}}\,v_{\Delta}\;.
\label{mnuII}
\end{equation}
%
Here  $h_{\ell\ell^\prime}$ is the matrix of Yukawa
couplings, which is directly related
to the PMNS neutrino mixing matrix $U_{\rm PMNS}\equiv U$:
\begin{equation}
h_{\ell\ell^\prime}\;\equiv\; \frac{1}{2v_\Delta}\left(U^*\,
\diag(m_1,m_2,m_3)\,U^\dagger\right)_{\ell\ell^\prime}\,.
\label{hU}
\end{equation}
%
It follows from the current data
on the parameter $\rho=M^2_W/M_Z^2\cos^2\theta_W$
that 
(see, e.g., \cite{vDrho})
$v_{\Delta}/v\leq 0.03$, or $v_{\Delta}<5$ GeV,
$v=174$ GeV being the SM Higgs doublet v.e.v.
We will consider in what follows values of
$v_{\Delta}$  lying roughly in the interval
$v_{\Delta} \sim (1 - 100)$ eV.
For $M_{\Delta} \sim (100 - 1000)$ GeV and
the indicated values of $v_{\Delta}$,
the rates of LFV processes involving
the $\mu^{\pm}$ can have values close
to the existing upper limits
(see \cite{Dinh:2012bp} and references
quoted therein).
A small value of  $v_{\Delta}$ implies that
that $\mu_{\Delta}$ has also to be small:
for $M_{\Delta} \sim v = 174$ GeV we have
$v_{\Delta} \cong \mu_{\Delta}$, while if
$M^2_{\Delta} >> v^2$, then
$v_{\Delta} \cong \mu_{\Delta} v^2/(2M^2_{\Delta})$
(see, e.g., \cite{Akeroyd:2009nu,vDrho}).
The requisite small value of $\mu_{\Delta}$, and thus of
 $v_{\Delta}$, can be generated, e.g., at higher orders
in perturbation theory~\cite{Chun:2003ej}
or in the context of theories with
extra dimensions~(see, e.g., \cite{Chen:2005mz}).

  The physical singly-charged Higgs scalar field
 practically coincides with the triplet scalar
field $\Delta^{+}$, the admixture of the doublet charged
scalar field being suppressed by the
factor $v_{\Delta}/v$. The singly- and doubly-  charged
Higgs scalars $\Delta^{+}$ and $\Delta^{++}$ have,
in general, different masses \cite{Chun:2003ej,HiggsMass}:
$m_{\Delta^{+}}\neq m_{\Delta^{++}}$.
Both possibilities $m_{\Delta^{+}} >  m_{\Delta^{++}}$ and
$m_{\Delta^{+}} < m_{\Delta^{++}}$ are allowed.
In what follows, for simplicity,
we will present numerical results for
$m_{\Delta^{+}} \cong m_{\Delta^{++}} \equiv M_{\Delta}$.

%
\mathversion{bold}
\subsection{The $\tau\rightarrow \mu\gamma$ and
$\tau\rightarrow e\gamma$ Decays}
\mathversion{normal}
%

In the Higgs triplet model considered, the
$\ell \to \ell^{\prime} + \gamma$ decay amplitude
receives at leading order contributions
from one loop diagrams with exchange of
virtual singly and doubly-charged Higgs scalars.
A detailed calculation of these contributions leads to the result
\cite{STPZee82,KOS_PLB566:2003,AAS_PRD79:2009,Dinh:2012bp}:
\begin{eqnarray}
{\rm BR}(\ell\to \ell^{\prime} + \gamma) =\; \frac{\alpha_{\rm em}}{192\,\pi}\,
\frac{\left|\left(h^{\dagger}h\right)_{\ell \ell^{\prime}}\right|^{2}}
{G_{F}^{2}}\,\left(
\frac{1}{m^2_{\Delta^{+}}} +
\frac{8}{m^2_{\Delta^{++}}}\right)^{2}
{\rm BR}(\ell \rightarrow \nu_\ell\, \ell^{\prime}\,
\overline{\nu}_{\ell^{\prime}})\,,
\label{muegII}
\end{eqnarray}
%
where $\ell = \mu$ and $\ell^{\prime}= e$, or $\ell = \tau$ and
$\ell^{\prime}=\mu,e$.
For $m_{\Delta^+}\approx m_{\Delta^{++}}=M_{\Delta}$,
the expression in eq. (\ref{muegII}) can be cast in the form:
\begin{eqnarray}
{\rm BR}(\ell \to \ell^{\prime} + \gamma) =
\; \frac{27\alpha_{\rm em}}{64\,\pi}\,
\frac{\left|\left(m^{\dagger}m\right)_{\ell \ell^{\prime}}\right|^{2}}
{16v_{\Delta}^4\,G_{F}^{2}\,M_{\Delta}^4}\,
{\rm BR}(\ell \rightarrow \nu_\ell\, \ell^{\prime}\,
\overline{\nu}_{\ell^{\prime}})\,.
\label{muegII2}
\end{eqnarray}
%
The factor $|(m^{\dagger}m)_{\ell \ell^{\prime}}|$, as it is not
difficult to show, is given by:
\begin{equation}
|\left(m^{\dagger}\,m\right)_{\ell \ell^{\prime}}| =
|U_{\ell 2}U^*_{\ell^{\prime} 2}\Delta m^2_{21}
+ U_{\ell 3}U^*_{\ell^{\prime} 3}\Delta m^2_{31}|\,,
\label{mdagml1l2}
\end{equation}
%
where we have used eqs. (\ref{mnuII}) and (\ref{hU})
and the unitarity of the PMNS matrix.
The expression in eq. (\ref{mdagml1l2}) is exact.
Obviously, $|(m^{\dagger}\,m)_{\ell \ell^{\prime}}|$
does not depend on the Majorana phases present in the
PMNS matrix $U$.

   The branching ratios, ${\rm BR}(\ell \to \ell^{\prime} + \gamma)$,
are inversely proportional to $(v_\Delta  M_\Delta)^4$. From the
the current upper bound on  ${\rm BR}(\mu \to e + \gamma)$,
eq. (\ref{mutoegexp}), and the expression for
$|(m^{\dagger}\,m)_{\mu e}|$ in terms of the neutrino oscillation
parameters, one can obtain a lower limit on $v_\Delta  M_\Delta$
\cite{Dinh:2012bp}:
\begin{equation}
v_{\Delta} > 2.98\times 10^2\,\left |
 s_{13}\,s_{23}\, \Delta m^2_{31} \right|^{\frac{1}{2}}\,
\left(\frac{100\,{\rm GeV}}{M_{\Delta}}\right)\,.
\label{vDmueg1}
\end{equation}
%
Using the the best fit values ($3\sigma$ allowed ranges) of
$\sin\theta_{13}$, $\sin\theta_{23}$ and $\Delta m^2_{31}$,
obtained in the global analysis \cite{Fogli:2012XY} we find:
\begin{equation}
v_{\Delta}\,M_{\Delta} > 4.60~(3.77)\times 10^{-7}~{\rm GeV^2}\,.
\label{vDmueg2}
\end{equation}
%

 As in the case of type I seesaw model, we can obtain an upper bounds
on the branching ratios  ${\rm BR}(\tau \to \mu + \gamma)$
and  ${\rm BR}(\tau \to e + \gamma)$
of interest using their relation with
${\rm BR}(\mu \to e + \gamma)$ and the current
experimental upper bound on ${\rm BR}(\mu \to e + \gamma)$.
We have:
\begin{equation}
\frac{{\rm BR}(\tau \to \mu (e) + \gamma)}
{{\rm BR}(\mu \to e + \gamma)} =
 \frac{\left|(m^+m)_{\tau\mu(e)}\right|^2}{\left|(m^+m)_{\mu e}\right|^2}\,
{\rm BR}(\tau \rightarrow \nu_\tau\,\mu (e)\,\bar{\nu}_{\mu(e)})\,.
\label{Rtmuegmueg}
\end{equation}
%
Using again the expressions for  $|(m^{\dagger}\,m)_{\ell \ell^{\prime}}|$
in terms of neutrino oscillation parameters
and the best fit values quoted in eqs. (\ref{Deltam2osc}) and
(\ref{thetaosc}) we get
in the case of NO (IO) neutrino mass spectrum:
\begin{equation}
4.41~(4.47) \leq \frac{\left|(m^+m)_{\tau\mu}\right|}
{\left|(m^+m)_{\mu e}\right|} \leq 5.57~(5.64)\,,~~{\rm NO~(IO)~~b.f.}
\label{Rtmugmuegbf}
\end{equation}
\begin{equation}
1.05~(1.03) \leq \frac{\left|(m^+m)_{\tau e}\right|}
{\left|(m^+m)_{\mu e}\right|} \leq 1.53~(1.51)~~~{\rm NO~(IO)~~b.f.}
\label{Rtegmuegbf}
\end{equation}
%
Employing the
$3\sigma$ allowed ranges of the
neutrino oscillation parameters derived in \cite{Fogli:2012XY}
we obtain:
\begin{equation}
0.87~(0.57)\leq  \frac{\left|(m^+m)_{\tau e}\right|}
{\left|(m^+m)_{\mu e}\right|} \leq 1.79~(1.78)~~~{\rm NO~(IO)~~2\sigma\,;}
\label{Rtsgmuegbf}
\end{equation}
\begin{equation}
3.07~(3.04)\leq  \frac{\left|(m^+m)_{\tau\mu}\right|}
{\left|(m^+m)_{\mu e}\right|}\leq 7.72~(7.85)~~~{\rm NO~(IO)~~3\sigma\,;}
\label{Rtmugmueg3s}
\end{equation}
\begin{equation}
0.55~(0.52)\leq \frac{\left|(m^+m)_{\tau e}\right|}
{\left|(m^+m)_{\mu e}\right|} \leq 1.95~(1.95)~~~{\rm NO~(IO)~~3\sigma\,.}
\label{Rtegmueg3s}
\end{equation}
%
From eqs. (\ref{mutoegexp}), (\ref{Rtmuegmueg}), (\ref{Rtmugmueg3s})
and  (\ref{Rtegmueg3s}) it follows that
\begin{equation}
{\rm BR}(\tau \to \mu + \gamma) < 5.9~(6.1)\times 10^{-12}\,,~~
{\rm BR}(\tau \to e + \gamma) < 3.9\times 10^{-13}\,,~~{\rm NO~(IO)}\,.
\label{tmugteglim}
\end{equation}
%
These values are significantly below the planned sensitivity of
the future experiments on the $\tau \to \mu + \gamma$ and
$\tau \to e + \gamma)$ decays. The observation of the any of the two
decays having a branching ratio definitely larger than that quoted in
eq. (\ref{tmugteglim}) would rule out the
TeV scale Higgs triplet model under discussion.
\begin{figure}
\begin{center}
\begin{tabular}{cc}
\includegraphics[width=7.5cm,height=6.5cm]{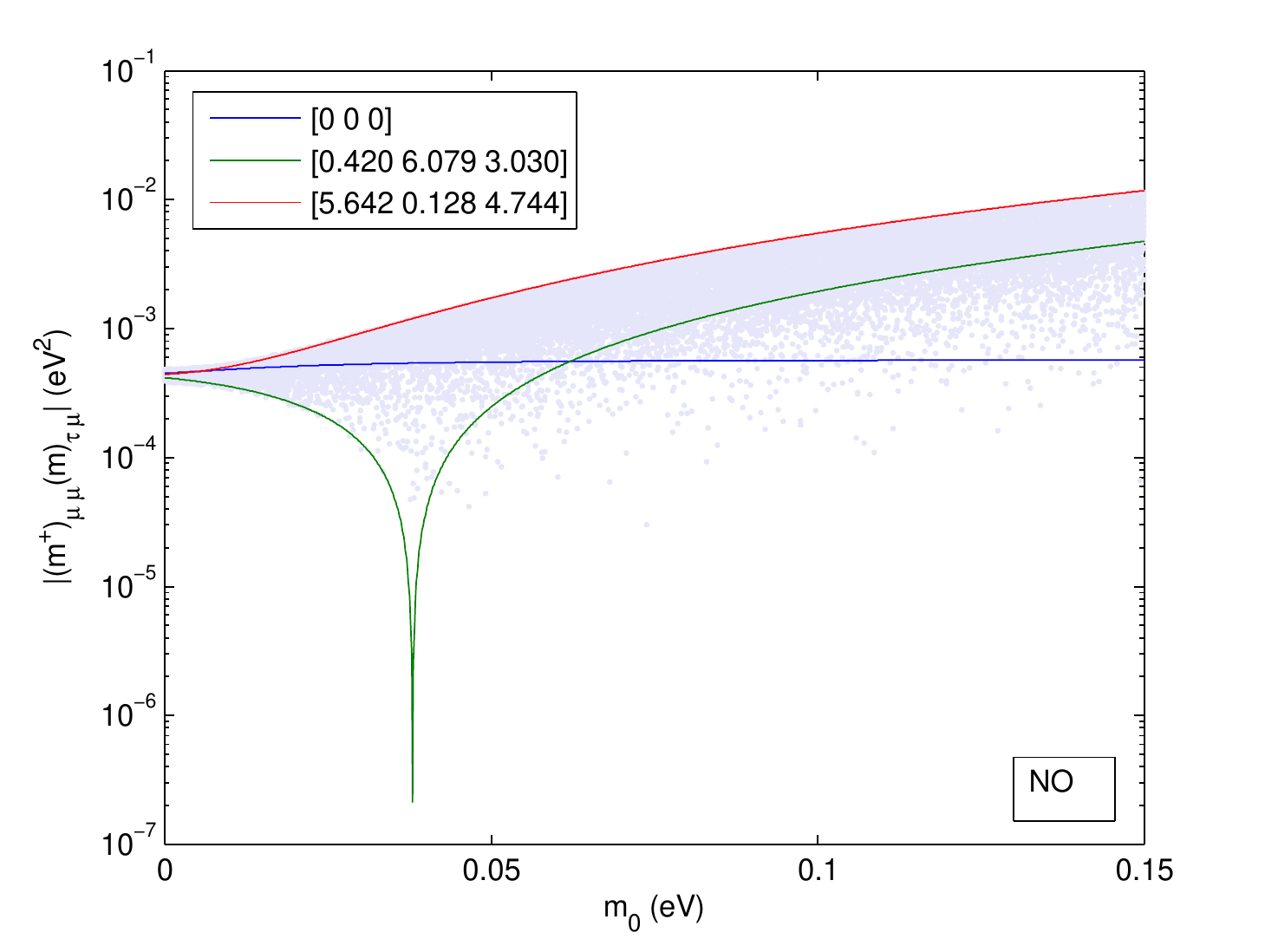} &
\includegraphics[width=7.5cm,height=6.5cm]{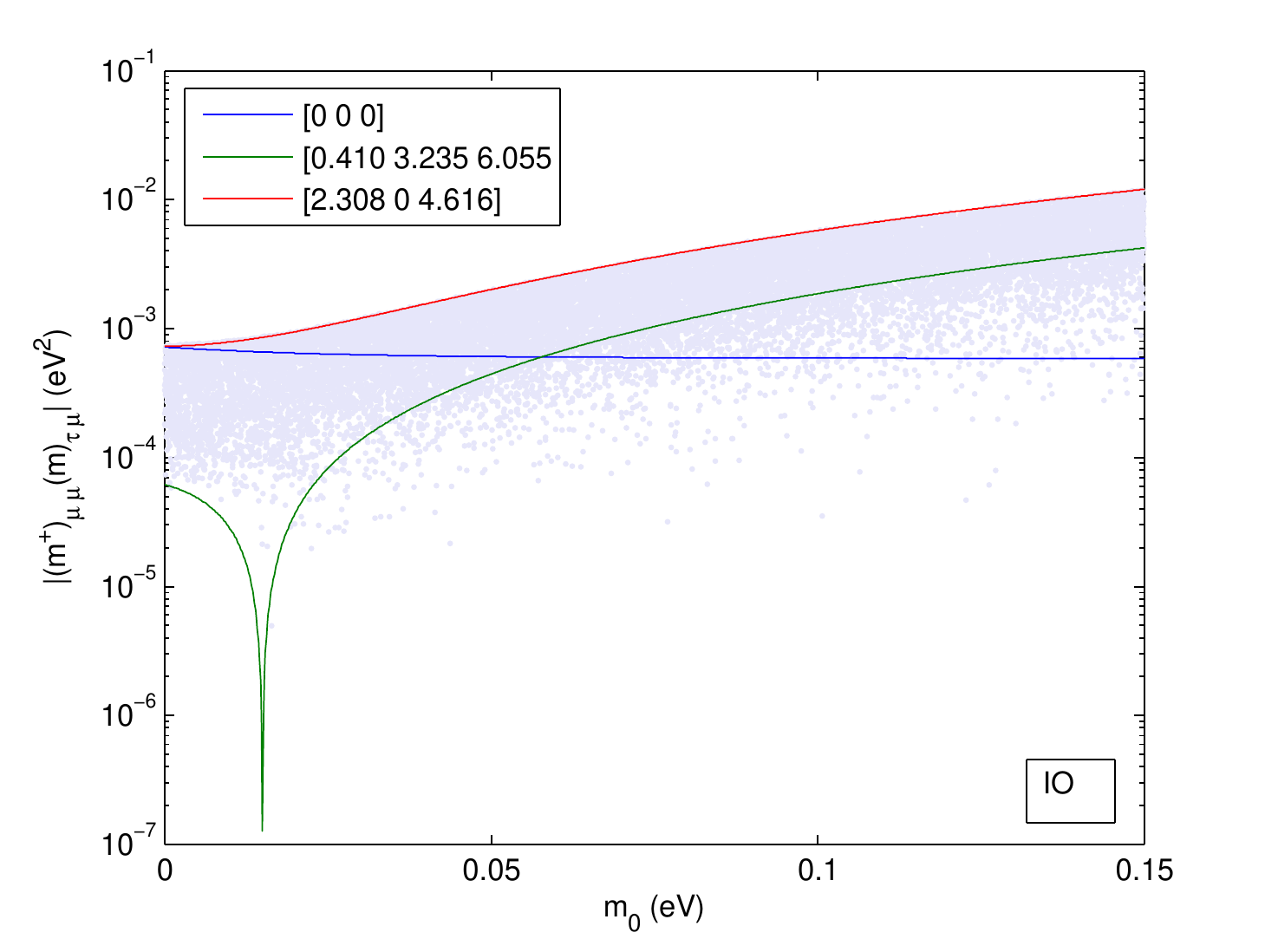}\\
\end{tabular}
\caption{The dependence of $|m_{\mu\mu}^*m_{\tau\mu}|$ on the lightest neutrino
mass $m_0$ in the cases of NO (left panel) and IO (right panel)
neutrino mass spectra, for three sets of values of
the Dirac and Majorana CPV phases, $[\delta, \alpha_{21}, \alpha_{31}]$.
The neutrino oscillation parameters $\sin\theta_{12}$, $\sin\theta_{23}$,
$\sin\theta_{13}$, $\Delta m^2_{21}$ and $\Delta m^2_{31}$
have been set to their best fit values,
eqs. (\ref{Deltam2osc}) and (\ref{thetaosc}).
The scattered points are obtained by varying
Dirac and Majorana CPV phases randomly
in the interval $[0,2\pi]$.
}
\label{Tau3Mu}
\end{center}
\end{figure}

%
\mathversion{bold}
\subsection{The $\tau\rightarrow 3\mu$ Decay}
\mathversion{normal}
%

 The leading contribution in the $\tau\to 3\mu$ decay amplitude
in the TeV scale HTM is due to a tree level diagram with exchange
of the virtual doubly-charged Higgs scalar $\Delta^{++}$.
The corresponding $\tau\to 3\mu$ decay branching ratio
is given by \cite{JBerna86} (see also, e.g.,
\cite{BiPet87,AAS_PRD79:2009}):
\begin{eqnarray}
{\rm BR}(\tau\to 3\mu) =\; \frac{\left|h^{*}_{\mu\mu}h_{\tau\mu}\right|^{2}}{G_{F}^{2}\,M_{\Delta}^4}\,
{\rm BR}(\tau\to\mu\bar{\nu}_\mu\nu_\tau)\,
=\; \frac{1}{G_{F}^{2}\,M_{\Delta}^4}\,
\frac{\left|m^{*}_{\mu\mu}m_{\tau\mu}\right|^{2}}
{16v_{\Delta}^4}{\rm BR}(\tau\to\mu\bar{\nu}_\mu\nu_\tau)\,,
\label{muegII2}
\end{eqnarray}
%
where $M_\Delta \equiv m_{\Delta^{++}}$ is the $\Delta^{++}$ mass
and we have neglected corrections $\sim m_\mu/m_\tau \cong 0.06$.

 Using the current upper bound on ${\rm BR(\tau\to 3\mu)}$,
eq. (\ref{tauto3muexp}),
and eq. (\ref{muegII2}),
we get the following constraint:
\begin{equation}
\left|h^{*}_{\mu\mu}h_{\tau\mu}\right| <
4.1\times10^{-5}\left(\frac{M_\Delta}{100~{\rm GeV}}\right)^2.
\end{equation}
%
Further, the lower limit on the product of $v_{\Delta}$ and
$M_\Delta$, eq. (\ref{vDmueg2}), implies the following upper limit on
${\rm BR(\tau\to 3\mu)}$:
 \begin{equation}
 {\rm BR(\tau\to 3\mu)} < 1.88~(4.17)\times 10^{-3}\,
 \frac{\left|m^{*}_{\mu\mu}m_{\tau\mu}\right|^{2}}
 {\left (1\,{\rm eV}\right)^4}\,.
 \label{mumutaumu1}
\end{equation}
%

The factor $|m^{*}_{\mu\mu}m_{\tau\mu}|$, as can be shown using
eqs. (\ref{mnuII}) and (\ref{hU}), depends not only on the
neutrino oscillation parameters,
but also on the type of the neutrino mass spectrum, the
lightest neutrino mass $m_0 \equiv {\rm min}(m_j)$, $j=1,2,3$
(i.e., on the absolute neutrino mass scale), and on the Majorana
CPV phases $\alpha_{21}$ and $\alpha_{31}$, present in the PMNS matrix.
The dependence of  $|m^{*}_{\mu\mu}m_{\tau\mu}|$ on $m_0$
for three sets of values of the CPV Dirac and Majorana phases $\delta$,
$\alpha_{21}$ and $\alpha_{31}$ in the cases of NO and IO
neutrino mass spectra is illustrated in Fig. \ref{Tau3Mu}.
The neutrino oscillation parameters were set to their best fit
values quoted in eqs. (\ref{Deltam2osc}) and (\ref{thetaosc}).
As Fig. \ref{Tau3Mu} indicates, both for the NO and IO spectra,
the maximal allowed value of $|m^{*}_{\mu\mu}m_{\tau\mu}|$
is a monotonically increasing function of $m_0$.

  The intervals of possible values of $|m^{*}_{\mu\mu}m_{\tau\mu}|$
in the cases of NO and IO neutrino mass spectra
determine the ranges of allowed values of
${\rm BR(\tau\to 3\mu)}$ in the TeV scale HTM.
Varying the three CPV phases independently in the interval
$[0,2\pi]$ and using the best fit, the $2\sigma$ and the $3\sigma$
allowed ranges of values of $\sin\theta_{12}$, $\sin\theta_{23}$,
$\sin\theta_{13}$, $\Delta m^2_{21}$ and  $\Delta m^2_{31}$
derived in \cite{Fogli:2012XY}, we get for $m_0 = 0;~0.01~;0,10$ eV:

\begin{itemize}

\item {$m_0=0$} eV, NO~(IO)

\begin{equation}
38.0~(5.35)\times 10^{-5}~{\rm eV^2}\leq\left|(m^*)_{\mu\mu}(m)_{\tau\mu}\right|\leq
4.82~(7.38)\times 10^{-4}~{\rm eV^2}~~~{\rm b.f};
\label{0mumutaumubf}
\end{equation}
\begin{equation}
2.77~(0.00)\times 10^{-4}~{\rm eV^2}\leq\left|(m^*)_{\mu\mu}(m)_{\tau\mu}\right|\leq
5.89~(8.11)\times 10^{-4}~{\rm eV^2}~~~{\rm 2\sigma};
\end{equation}
\begin{equation}
2.33~(0.00)\times 10^{-4}~{\rm eV^2}\leq\left|(m^*)_{\mu\mu}(m)_{\tau\mu}\right|\leq
8.35~(8.45)\times 10^{-4}~{\rm eV^2}~~~{\rm 3\sigma}.
\label{0mumutaumu3s}
\end{equation}

\item {$m_0=0.01$} eV, NO~(IO)

\begin{equation}
33.6~(1.66)\times 10^{-5}~{\rm eV^2}\leq\left|(m^*)_{\mu\mu}(m)_{\tau\mu}\right|\leq
5.34~(8.06)\times 10^{-4}~{\rm eV^2}~~~{\rm b.f};
\end{equation}
\begin{equation}
2.24~(0.00)\times 10^{-4}~{\rm eV^2}\leq\left|(m^*)_{\mu\mu}(m)_{\tau\mu}\right|\leq
6.41~(8.99)\times 10^{-4}~{\rm eV^2}~~~{\rm 2\sigma};
\end{equation}
\begin{equation}
1.76~(0.00)\times 10^{-4}~{\rm eV^2}\leq\left|(m^*)_{\mu\mu}(m)_{\tau\mu}\right|\leq
8.96~(9.41)\times 10^{-4}~{\rm eV^2}~~~{\rm 3\sigma}.
\label{001mumutaumu3s}
\end{equation}

\item {$m_0=0.1$} eV, NO~(IO)

\begin{equation}
0.00~(0.00)~{\rm eV^2}\leq\left|(m^*)_{\mu\mu}(m)_{\tau\mu}\right|\leq
5.48~(5.76)\times 10^{-3}~{\rm eV^2}~~~{\rm b.f};
\end{equation}
\begin{equation}
0.00~(0.00)~{\rm eV^2}\leq\left|(m^*)_{\mu\mu}(m)_{\tau\mu}\right|\leq
5.57~(5.85)\times 10^{-3}~{\rm eV^2}~~~{\rm 2\sigma};
\end{equation}
\begin{equation}
0.00~(0.00)~{\rm eV^2}\leq\left|(m^*)_{\mu\mu}(m)_{\tau\mu}\right|\leq
5.85~(5.88)\times 10^{-3}~{\rm eV^2}~~~{\rm 3\sigma}.
\label{01mumutaumu3s}
\end{equation}

\end{itemize}

 We would like to determine next whether
${\rm BR(\tau\to 3\mu)}$ predicted by the TeV scale
HTM considered can be bigger than the
sensitivity limit of $\sim 10^{-10}$ of the future
planned experiment on $\tau\to 3\mu$ decay,
given the stringent upper bounds on the $\mu\to e +\gamma$
and  $\mu\to 3e$ decay branching ratios,
eqs. (\ref{mutoegexp}) and (\ref{muto3eexp}).
As we have seen, the current upper bound
on  ${\rm BR(\mu\to e+\gamma)}$ leads to the
lower limit eq. (\ref{vDmueg2}) of $v_{\Delta}M_{\Delta}$.
We have to take into account also the important
constraint on ${\rm BR(\tau\to 3\mu)}$
following from the current upper bound on $\mu\to 3e$ decay
branching ratio ${\rm BR(\mu\to 3e)}$, eq. (\ref{muto3eexp}).
In the case of ${\rm BR(\mu\to 3e)}$ we have
${\rm BR(\mu\to 3e)} \propto |m^{*}_{\mu e}m_{e e}|^2$.
The quantity $|m^{*}_{\mu e}m_{e e}|$, and thus
${\rm BR(\mu\to 3e)}$, depends on the same set of
neutrino mass and mixing parameters as
$|(m^*)_{\mu\mu}(m)_{\tau\mu}|$, and thus ${\rm BR(\tau\to 3\mu)}$.
We have performed a numerical analysis in order to determine
the regions of values of the neutrino oscillation parameters
and of the three CPV phases $\delta$, $\alpha_{21}$ and
$\alpha_{31}$, in which the experimental
upper bounds on  ${\rm BR(\mu\to e+\gamma)}$ and ${\rm BR(\mu\to 3e)}$,
eqs. (\ref{mutoegexp}) and (\ref{muto3eexp}),
and the following requirement,
\begin{equation}
 10^{-10} \leq {\rm BR(\tau\to 3\mu)} \leq 10^{-8}\,,
\label{t3mRange}
\end{equation}
%
%
\begin{figure}[t]
\begin{center}
\begin{tabular}{cc}
\includegraphics[width=6cm,height=5.5cm]{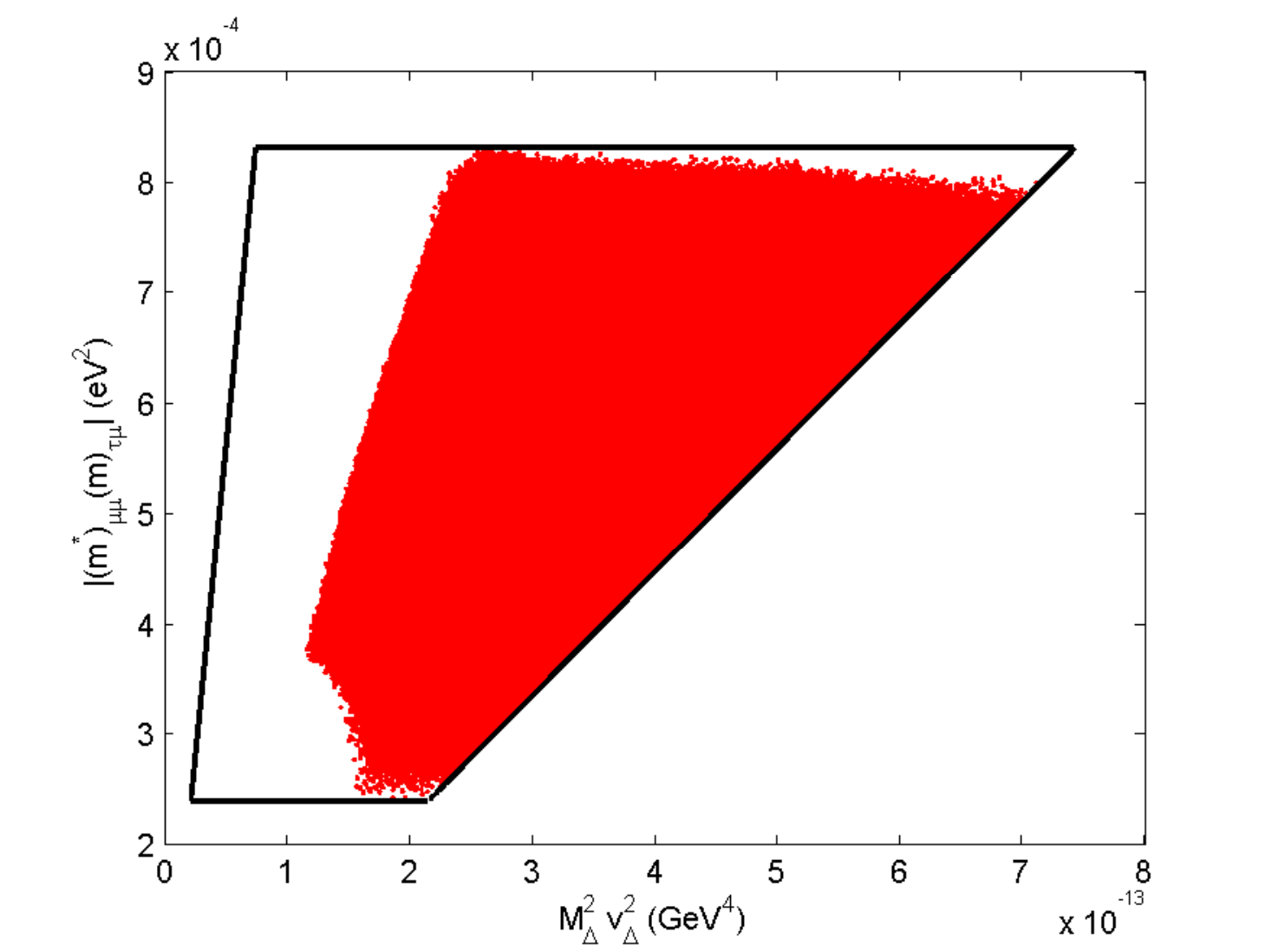} &
\includegraphics[width=6cm,height=5.5cm]{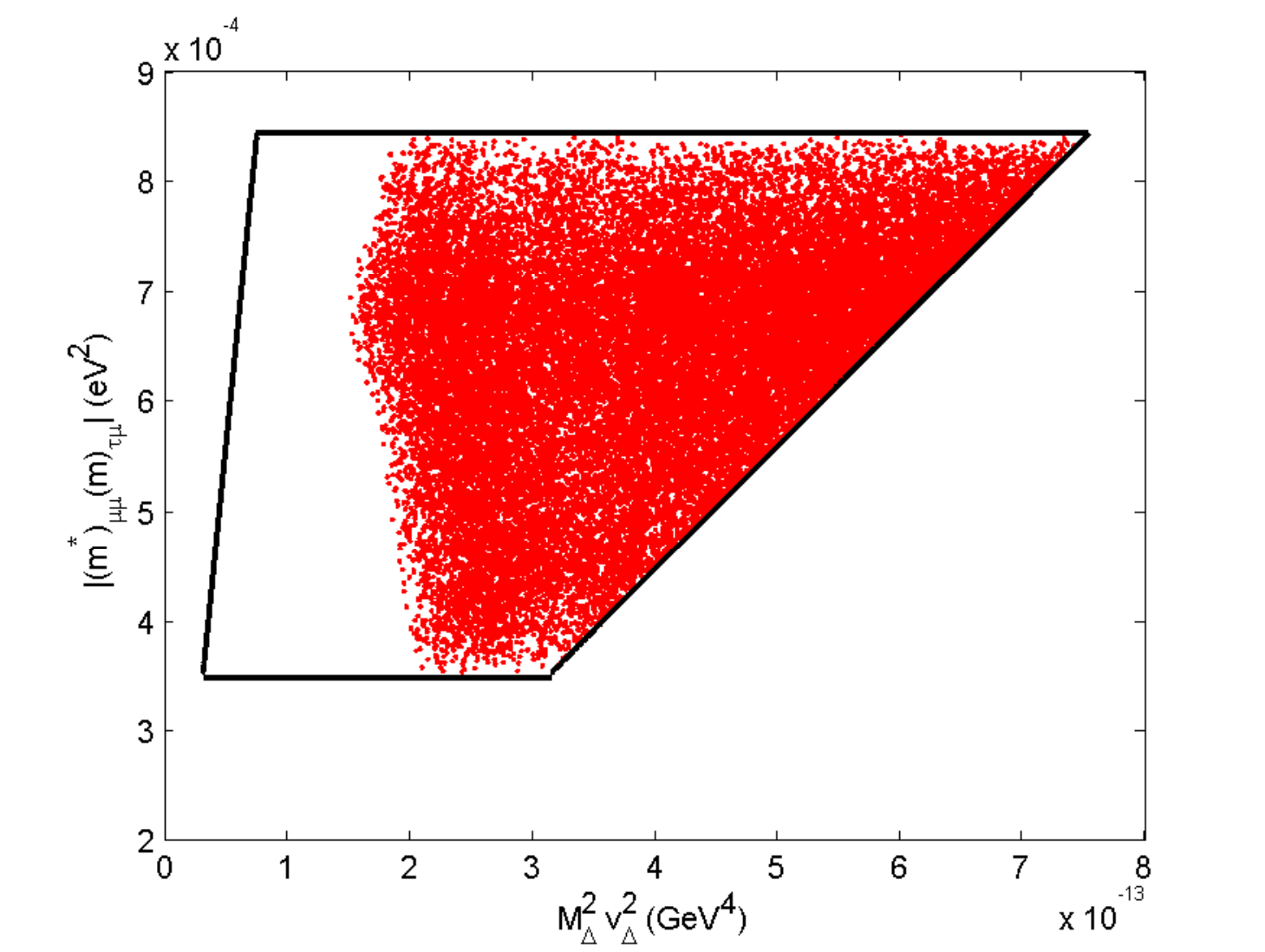}\\
\includegraphics[width=6cm,height=5.5cm]{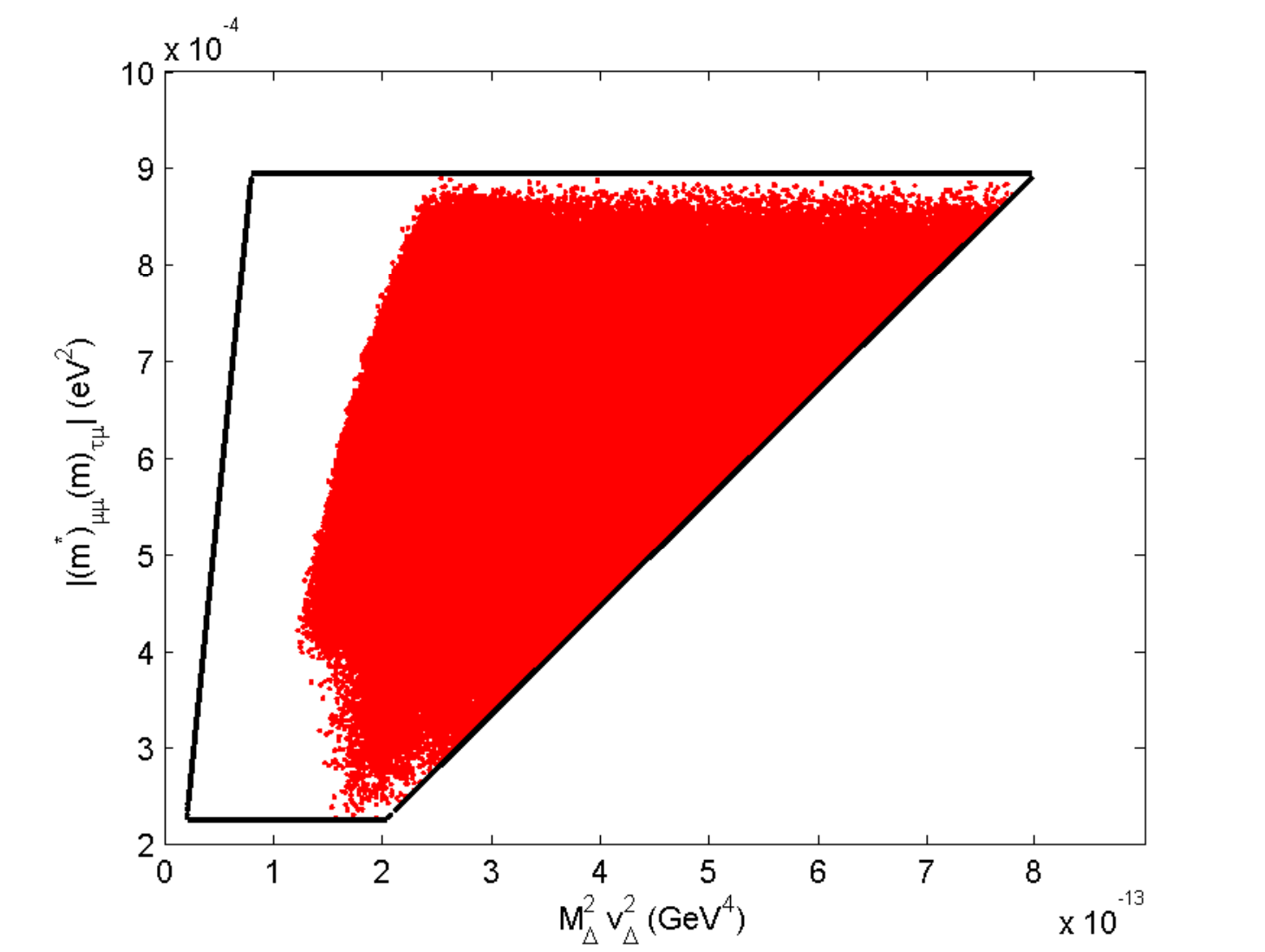} &
\includegraphics[width=6cm,height=5.5cm]{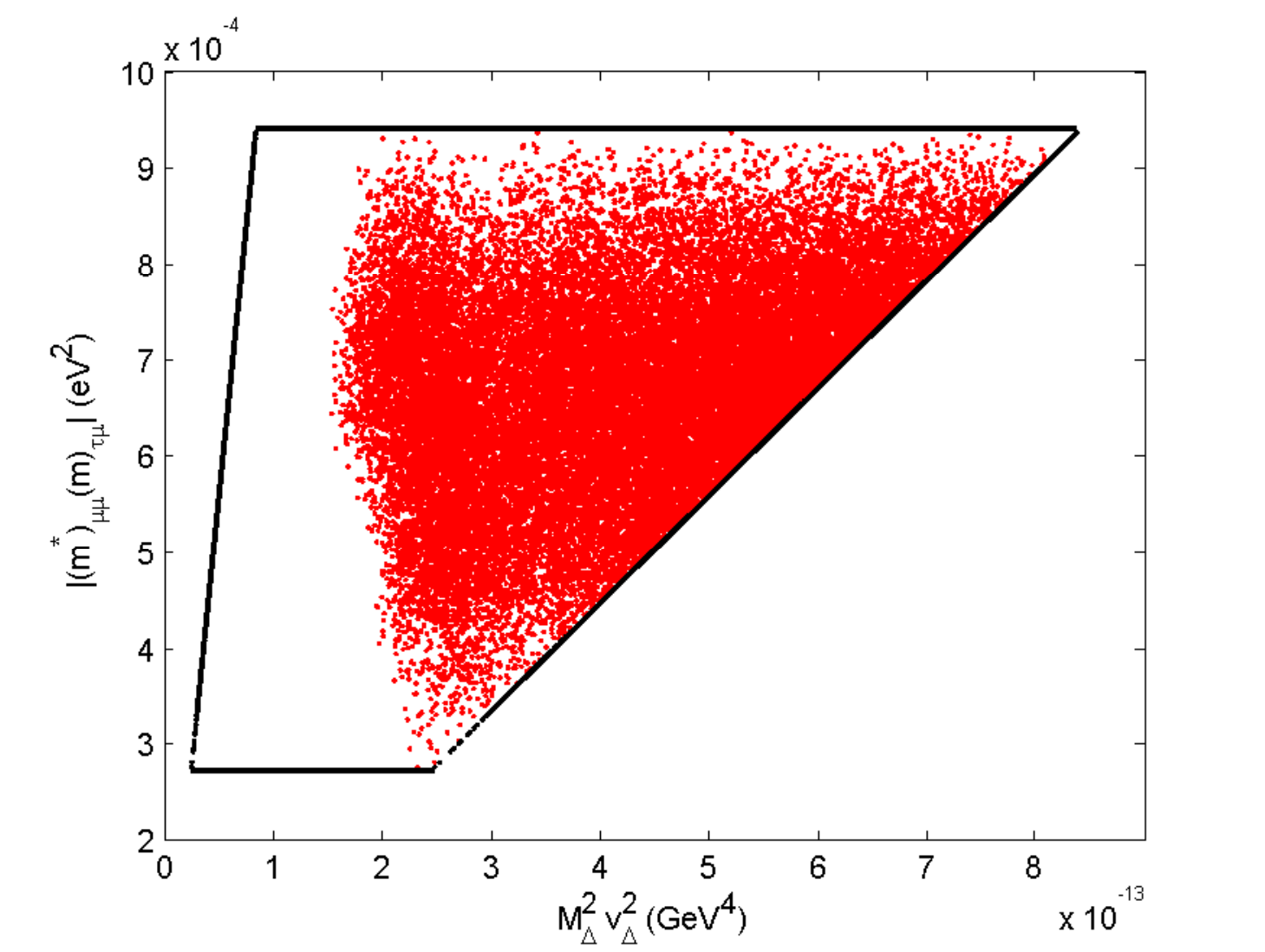}\\
\includegraphics[width=6cm,height=5.5cm]{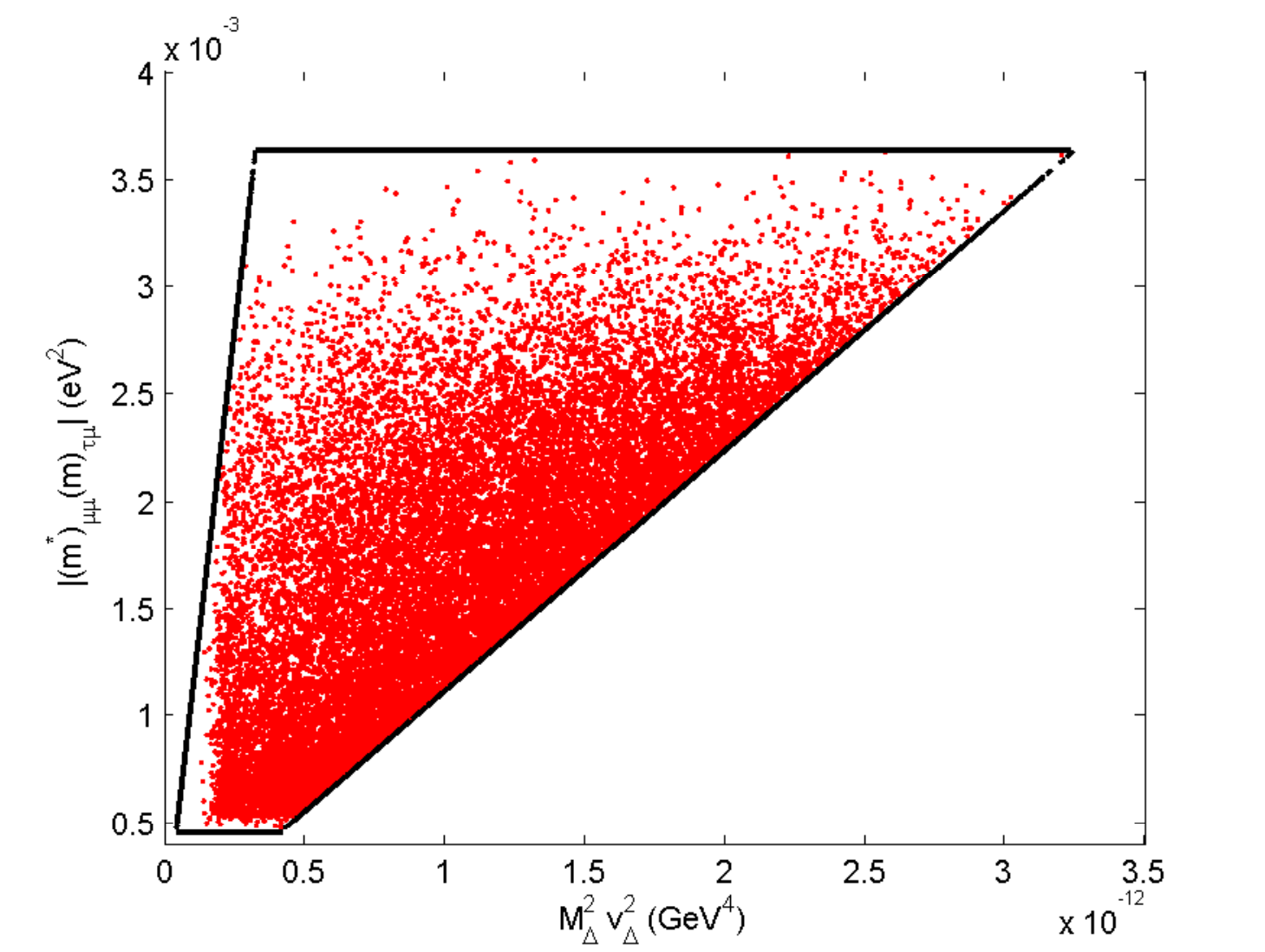} &
\includegraphics[width=6cm,height=5.5cm]{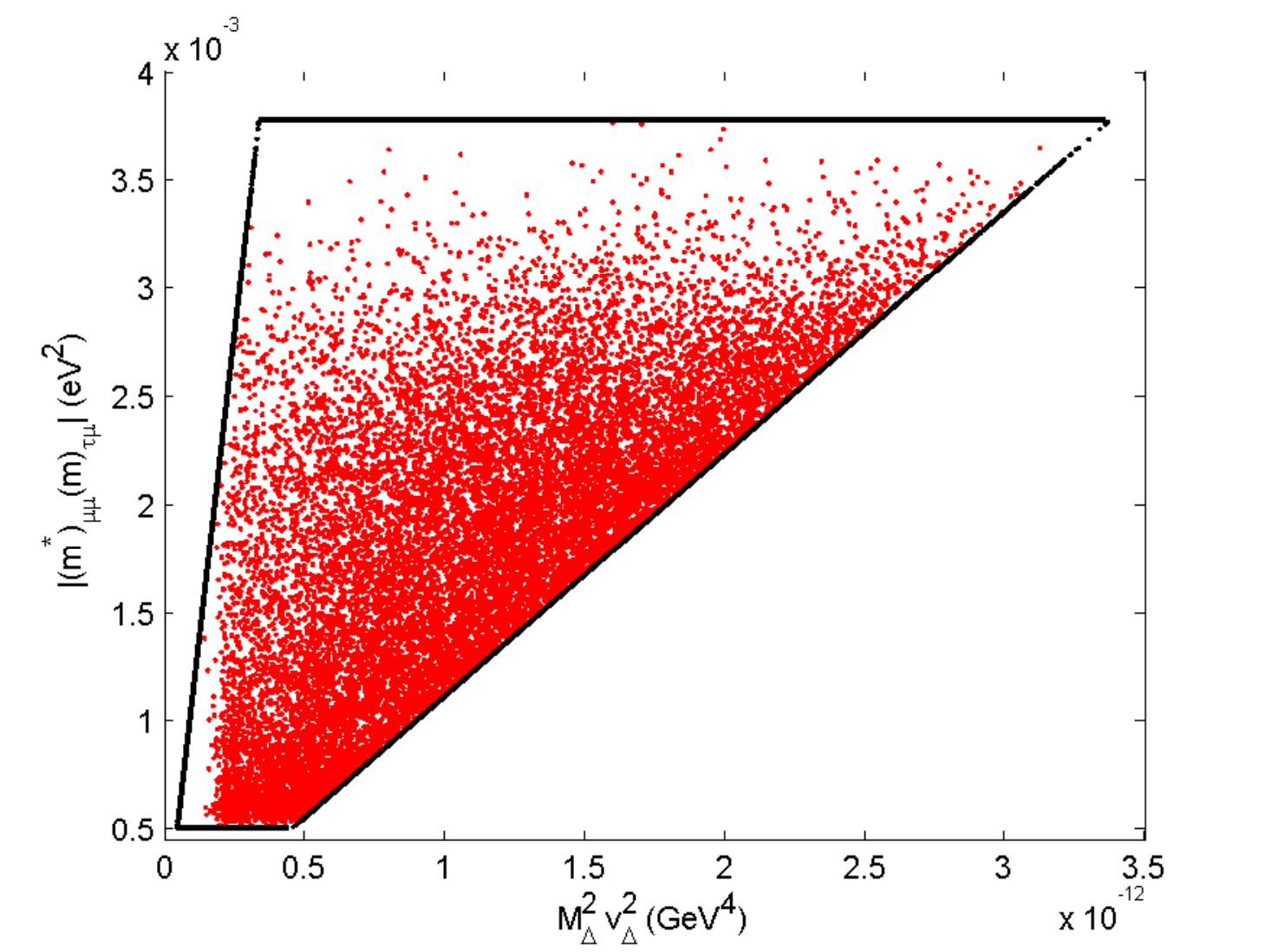}
\end{tabular}
\caption{The regions in the  $v^2_{\Delta}M_\Delta^2 -
|(m^*_{\mu\mu})(m_{\tau\mu})|$ plane
where $10^{-10}\leq BR(\tau\to 3\mu) \leq 10^{-8}$
(the areas deliminated by the black lines)
and the
the upper limits $BR(\tau\to 3e) <10^{-12}$ and
$BR(\tau\to e\gamma) <5.7\times 10^{-13}$ are
satisfied (the colored areas),
for  $m_0=0$ (upper panels), 0.01 eV (middle panels),
0.10 eV (lower panels) and NO (left panels) and IO (right panels)
neutrino mass spectra. The figures are obtained by varying
the neutrino oscillation parameters in
their $3\sigma$ allowed ranges \cite{Fogli:2012XY};
the CPV Dirac and Majorana phases were varied
in the interval $[0,2\pi]$.
}
\label{M2V2Constraint}
\end{center}
\end{figure}
%
are simultaneously satisfied.
The analysis is performed for three values of
$m_0 = 0$; 0.01 eV; 0.10 eV.
The neutrino oscillation parameters
 $\sin\theta_{12}$, $\sin\theta_{23}$,
$\sin\theta_{13}$, $\Delta m^2_{21}$ and  $\Delta m^2_{31}$
were varied in their respective $3\sigma$ allowed ranges
taken from \cite{Fogli:2012XY}. The CPV phases
$\delta$, $\alpha_{21}$ and $\alpha_{31}$
were varied independently in the interval $[0,2\pi]$.
The results of this analysis are presented graphically
in Fig. \ref{M2V2Constraint}, in which we show the regions of values of the
quantities  $|m^{*}_{\mu\mu}m_{\tau\mu}|$ and
$v_{\Delta}M_{\Delta}$ where the three conditions
(\ref{mutoegexp}), (\ref{muto3eexp}) and (\ref{t3mRange})
are simultaneously fulfilled in the cases of
$m_0 = 0$; 0.01 eV; 0.10 eV for the NO and IO spectra.
For $m_0 = 0$ and NO spectrum, the results depend
weakly on the CPV phases; they are independent
of the phase $\alpha_{31}$ if  $m_0 = 0$ and
the spectrum is of the IO type.
The analysis performed by us shows that
the maximal values  ${\rm BR(\tau\to 3\mu)}$ can have
are the following:
\begin{eqnarray}
\label{0maxt3mu}
{\rm BR(\tau\to 3\mu)} \leq 1.02~(1.68)\times 10^{-9}\,,
~~~m_0=0~{\rm eV,~~NO~(IO)}\,,\\
\label{001maxt3mu}
{\rm BR(\tau\to 3\mu)} \leq 1.24~(2.05)\times 10^{-9}\,,
~~~m_0=0.01~{\rm eV,~~NO~(IO)}\,,\\
\label{01maxt3mu}
{\rm BR(\tau\to 3\mu)} \leq  8.64~(9.11)\times 10^{-9}\,,
~~~m_0=0.10~{\rm eV,~~NO~(IO)}\,.
\end{eqnarray}

Thus, for all the three values of
$m_0$ considered, which span essentially
the whole interval of possible values of $m_0$,
the maximal allowed values of
${\rm BR(\tau\to 3\mu)}$ is by
a factor of $\sim 10$ to $\sim 90$ bigger than
the projected sensitivity limit of $10^{-10}$ of
the future experiment on the $\tau \to 3\mu$ decay.
The regions on the
$|m^{*}_{\mu\mu}m_{\tau\mu}|- v_{\Delta}M_{\Delta}$
plane, where the three conditions of interest are satisfied, are sizeable.
The maximal value of ${\rm BR(\tau\to 3\mu)}$
for, e.g.,  $m_0=0.01$ eV and NO (IO) spectrum,
quoted in eq. (\ref{001maxt3mu}), is reached for
$\sin^2\theta_{12} = 0.269~(0.308)$,
$\sin^2\theta_{23} = 0.527~(0.438)$,
$\sin^2\theta_{13} = 0.0268~(0.0203)$,
$\Delta m^2_{21} = 7.38~(7.56)\times 10^{-5}~{\rm eV^2}$,
$\Delta m^2_{31} = 2.14~(2.40)\times 10^{-3}~{\rm eV^2}$
and $[\delta,\alpha_{21},\alpha_{31}] = [2.300,5.098,3.437]$~
([1.577,0.161,3.436]).

 As it follows from Fig. \ref{Tau3Mu} and the results quoted
in eqs. (\ref{0mumutaumubf}) - (\ref{01mumutaumu3s}),
for certain values of the absolute neutrino mass scale $m_0$
and the CPV phases, $|m^{*}_{\mu\mu}m_{\tau\mu}|$ can be strongly
suppressed; we can have even  $|m^{*}_{\mu\mu}m_{\tau\mu}| = 0$.
For NO (IO) neutrino mass spectrum, such a strong
suppression can happen for
$m_0 \gtap 38~{\rm meV}$ ($m_0 \gtap 15~{\rm meV}$).
The strong suppression of $|m^{*}_{\mu\mu}m_{\tau\mu}|$
seen in Fig. \ref{Tau3Mu} takes place in the case of NO (IO)
spectrum at $m_0 = 38~{\rm meV}$ and
$[\delta,\alpha_{21},\alpha_{31}] = [0.420,6.079,3.030]$
($m_0 = 15~{\rm meV}$ and
$[\delta,\alpha_{21},\alpha_{31}] = [0.410,3.235,6.055]$).
For $m_0=0.10~{\rm eV}$, for instance,
we have $|m^{*}_{\mu\mu}m_{\tau\mu}| = 0$ in the case of NO
mass spectrum at $\delta=2.633$, $\alpha_{21}=2.533$
and $\alpha_{31}=5.349$, while for the IO spectrum
$|m^{*}_{\mu\mu}m_{\tau\mu}|$ goes through zero for
$\delta=4.078$, $\alpha_{21}=2.161$ and $\alpha_{31}=5.212$.
The above examples of the vanishing of $|m_{\mu\mu}^*m_{\tau\mu}|$
when $m_0=0.10$ eV are not unique, it can happen also at other
specific sets of values of the Dirac and Majorana CPV phases.

  If in the planned experiment on the $\tau \to 3\mu$ decay
the limit  ${\rm BR(\tau\to 3\mu)} < 10^{-10}$ will be obtained,
this will imply the following upper limit on the product
$|h^{*}_{\mu\mu}h_{\tau\mu}|$ of Yukawa couplings:
\begin{equation}
\left|h^{*}_{\mu\mu}h_{\tau\mu}\right| <
2.83\times10^{-6}\left(\frac{M_\Delta}{100~{\rm GeV}}\right)^2.
\end{equation}

%
\section{Conclusions}
%

  In the present article we have investigated in
detail the $\tau\to (e,\mu) + \gamma$ and
$\tau\to 3\mu$ decays in the TeV scale
type I see-saw and Higgs Triplet models of
neutrino mass generation. Future experiments
at the SuperB factory are planned to have
sensitivity to the branching rations of the these decays
${\rm BR(\tau\to (e,\mu) + \gamma)} \gtap 10^{-9}$
and ${\rm BR(\tau\to 3\mu)} \gtap 10^{-10}$,
which is an improvement by one and two orders
of magnitude with respect to that reached so far
in the searches for the $\tau\to (e,\mu) + \gamma$
and $\tau\to 3\mu$ decays, respectively.
In the models we have considered the
scale of new physics associated with the existence of
nonzero neutrino masses and neutrino mixing is assumed
to be in the range of $\sim (100 - 1000)$ GeV.
In the type I see-saw scenario this scale is determined by the masses
of the heavy Majorana neutrinos, while in the Higgs Triplet
model it corresponds to the masses of the new singly charged,
doubly charged and neutral physical Higgs particles.
In the type I see-saw class of models of interest,
the flavour structure of the couplings of the
new particles - the heavy Majorana neutrinos $N_j$ -
to the charged leptons and $W^{\pm}$-boson
and to the flavour neutrino fields and the $Z^0$-boson,
$(RV)_{lj}$, $l=e,\mu,\tau$,
are basically determined by
the requirement of reproducing the data on the neutrino
oscillation parameters (see, e.g., \cite{Ibarra:2011xn}).
In the Higgs Triplet model the Yukawa couplings
of the new scalar particles to the charged leptons
and neutrinos are proportional
to the Majorana mass matrix of the
LH active flavour neutrinos. As a consequence, the rates of
the LFV processes in the charged lepton sector
can be calculated in both models
in terms of a few unknown parameters.
These parameters are constrained by different sets of
data such as, e.g., data on neutrino oscillations,
from EW precision tests, on the LFV violating processes
$\mu\rightarrow e+\gamma$, $\mu\rightarrow 3e$, etc.
In the TeV scale type I see-saw  scenario considered
all the constraints can be satisfied
for sizeable values of the couplings $|(RV)_{lj}|$
in a model \cite{Ibarra:2011xn} with two heavy
Majorana neutrinos $N_{1,2}$, in which
the latter have close masses forming a pseudo-Dirac state,
$M_2 = M_1 (1 + z)$, $M_{1,2},z >0$, $z \ll 1$,
and their charged and neutral current couplings
(see eqs. (\ref{NCC}) and (\ref{NNC})),
$(RV)_{lj}$, $j=1,2$, satisfy eq. (\ref{rel0}).
In this scheme the lightest neutrino mass $m_0 = 0$
and the neutrino mass spectrum is either normal
hierarchical (NH) or inverted hierarchical (IH),

  We find using the constraints on the couplings
$(RV)_{lj}$, $j=1,2$, from the low energy
electroweak precision data,
eqs. (\ref{e-bound}) - (\ref{tau-bound}),
that the branching ratios of the decays
$\tau\to (e,\mu) + \gamma$ and $\tau\to 3\mu$
predicted in the TeV scale type I see-saw model can at
most be of the order of the sensitivity of the planned
future experiments,
${\rm BR(\tau\to (e,\mu) + \gamma)} \ltap 10^{-9}$
and ${\rm BR(\tau\to 3\mu)} \ltap 10^{-10}$.
Taking into account the
stringent experimental upper bounds on the
$\mu\to e + \gamma$ and $\mu\to 3e$ decay rates
has the effect of constraining further
the maximal values of ${\rm BR(\tau\to (e,\mu) + \gamma)}$
and ${\rm BR(\tau\to 3\mu)}$ compatible with the
data. In the case of NH spectrum, for instance,
we get using the $2\sigma$ ($3\sigma$) ranges of the neutrino
oscillations parameters from \cite{Fogli:2012XY} and
varying the CPV Dirac and Majorana phases
$\delta$, $\alpha_{21}$ and  $\alpha_{31}$
independently in the interval $[0,2\pi]$:
${\rm BR(\tau\to e + \gamma)} \ltap 1.4~(1.7)\times 10^{-12}$,
${\rm BR(\tau\to \mu + \gamma)} \ltap 4.9~(9.7)\times 10^{-11}$,
and ${\rm BR(\tau\to 3\mu)} \ltap 3.8~(7.4)\times 10^{-12}$.
For specific values of the neutrino mixing parameters
in the case of the IH spectrum, the predicted rates of the
$\mu\to e + \gamma$ and $\mu\to 3e$ decays
are strongly suppressed   and the experimental
upper bounds on these rates are automatically satisfied.
In this special case the $\tau\to \mu + \gamma$ and the
$\tau\to 3\mu$ decay rates are also predicted to
be strongly suppressed and significantly smaller
than the planned sensitivity of the future experiments,
while for the $\tau\to e + \gamma$ decay we have
${\rm BR(\tau\to e + \gamma)} \ltap 10^{-9}$.
Clearly, if any of the three $\tau$ decays under
discussion is observed in the planned experiments,
the TeV scale type I see-saw model we have considered
will be strongly disfavored if not ruled out.

 The predicted rates of the  $\mu\to e + \gamma$
and of the $\tau\to (e,\mu) + \gamma$ decays in the
Higgs Triplet model are also correlated.
Using the existing experimental upper bound on
${\rm BR(\mu\to e + \gamma)}$ we find the following
upper limits on the  $\tau\to \mu + \gamma$ and
$\tau\to e + \gamma$ decay branching ratios for the
NO (IO) neutrino mass spectrum:
${\rm BR(\tau\to \mu + \gamma)} \ltap 5.9~(6.1)\times 10^{-12}$,
${\rm BR(\tau\to e + \gamma)} \ltap 3.9\times 10^{-12}$.
These values are significantly below the planned sensitivity of
the future experiments on the $\tau \to \mu + \gamma$ and
$\tau \to e + \gamma$ decays. The observation of the any of the two
decays having a branching ratio definitely larger than that
quoted above would rule out the
TeV scale Higgs triplet model under discussion.
In contrast, we find that in a sizeable region of the parameter
space of the Higgs Triplet model, the $\tau\to 3\mu$ decay
branching ratio ${\rm BR(\tau\to 3\mu)}$ can have a value
in the interval $(10^{-10} - 10^{-8})$ and the predicted
values of ${\rm BR(\mu \to e + \gamma)}$
and ${\rm BR(\mu\to 3e)}$ satisfy the existing stringent
experimental upper bounds. Thus, the observation of
the $\tau\to 3\mu$ decay with
${\rm BR(\tau\to 3\mu)} \gtap 10^{-10}$
and the non-observation of the $\tau\to \mu + \gamma$ and
$\tau\to e + \gamma$ decays in the planned experiments having
a sensitivity to ${\rm BR(\tau\to (e,\mu) + \gamma)} \geq 10^{-9}$,
would constitute an evidence in favor of the Higgs Triplet model.

 To conclude, the planned searches for the
$\tau\to \mu + \gamma$, $\tau\to e + \gamma$
and $\tau\to 3\mu$ decays with sensitivity to
${\rm BR(\tau\to (e,\mu) + \gamma)} \gtap 10^{-9}$
and to ${\rm BR(\tau\to 3\mu)} \gtap 10^{-10}$
will provide additional important test
of the TeV scale see-saw type I and Higgs Triplet models
of neutrino mass generation.

\section*{Acknowledgments}
This work was supported in part by the INFN program on
``Astroparticle Physics'' (D.N.D. and S.T.P.),
by the World Premier International
Research Center Initiative (WPI Initiative), MEXT,
Japan,  and by the European Union FP7-ITN INVISIBLES
(Marie Curie Action, PITAN-GA-2011-289442) (S.T.P.).

\end{document}